\documentclass[reprint,showpacs,preprintnumbers,amsmath,amssymb,showkeys,pre]{revtex4}
\usepackage{graphicx}
\usepackage{dcolumn}
\usepackage{bm}
\usepackage{amsmath}
\usepackage{amssymb}
\usepackage{amsfonts}
\usepackage{color}
\usepackage{dsfont}
\usepackage{epsfig}
\usepackage{hyperref}
\usepackage{mathrsfs}
\usepackage{multirow}
\usepackage{pifont}
\usepackage[T1]{fontenc}
\usepackage{relsize}
\usepackage{graphicx}
\usepackage{wasysym}
\usepackage{txfonts}   
\usepackage{subeq}
\usepackage{slashbox}
\usepackage[normalem]{ulem}
\usepackage[usenames,dvipsnames]{xcolor}

\newcommand{\ds}{\displaystyle }

\newcommand{\sgn}{\text{sgn}}
\def \gv#1{\mbox{\boldmath $#1$}}
\renewcommand \vec \gv

\graphicspath{ {./Figures/} }

\begin{document}


\title{Mechanical energy dissipation induced by sloshing and wave breaking in a fully coupled angular motion system. Part I: Theoretical formulation and Numerical investigation.} 

\author{B. Bouscasse}
\email{benjamin.bouscasse@cnr.it}
\affiliation{CNR-INSEAN \\ \mbox{Marine Technology Research Institute, Rome, Italy}}
\affiliation{Aeronautics Department (ETSIA), \mbox{Technical University of Madrid (UPM), 28040 Madrid, Spain}}
\author{A. Colagrosssi}
\email{andrea.colagrossi@cnr.it}
\affiliation{CNR-INSEAN \\ \mbox{Marine Technology Research Institute, Rome, Italy}}
\author{A. Souto-Iglesias}
\email{antonio.souto@upm.es}
\affiliation{Naval Architecture Department (ETSIN), \mbox{Technical University of Madrid (UPM), 28040 Madrid, Spain}}
\author{J. L. Cercos-Pita}
\email{jl.cercos@upm.es}
\affiliation{Naval Architecture Department (ETSIN), \mbox{Technical University of Madrid (UPM), 28040 Madrid, Spain}}

\date{\today}

\begin{abstract}
A dynamical system involving a driven pendulum filled with liquid, is analyzed in the present paper series.
The study of such a system is conducted in order to understand energy dissipation resulting from the shallow water sloshing and induced wave breaking. This analysis is relevant for the design of Tuned Liquid Damper devices. The complexity and violence of the flow generated by the roll motion
results in the impossibility of using an analytical approach, requiring in turn the use of a suitable numerical solver. In Part I, the coupled dynamical system is thoroughly described, revealing its nonlinear features associated with the large amplitude of the forcing, both in terms of mechanical and fluid dynamical aspects. A smoothed particle hydrodynamics (SPH) model, largely validated in literature, is used to calculate the frequency behavior of the whole system. For small rotation angles, a semi-analytical model of the energy dissipated by the fluid, based on a hydraulic jump solution, is developed; the energy transfer is numerically calculated in order to extend the analysis to large oscillation angles.
The experimental part of the investigation is carried out in Part II of this work.
\end{abstract}


\maketitle

\section{Introduction}
The present work deals with energy dissipation induced by sloshing in a fully coupled angular motion system. The investigation is completed by the experimental study of Part II \citep{bouscasse2013mechanical_partII_ARXIV}.

\begin{figure}[h!]
\includegraphics[width=0.95\textwidth]{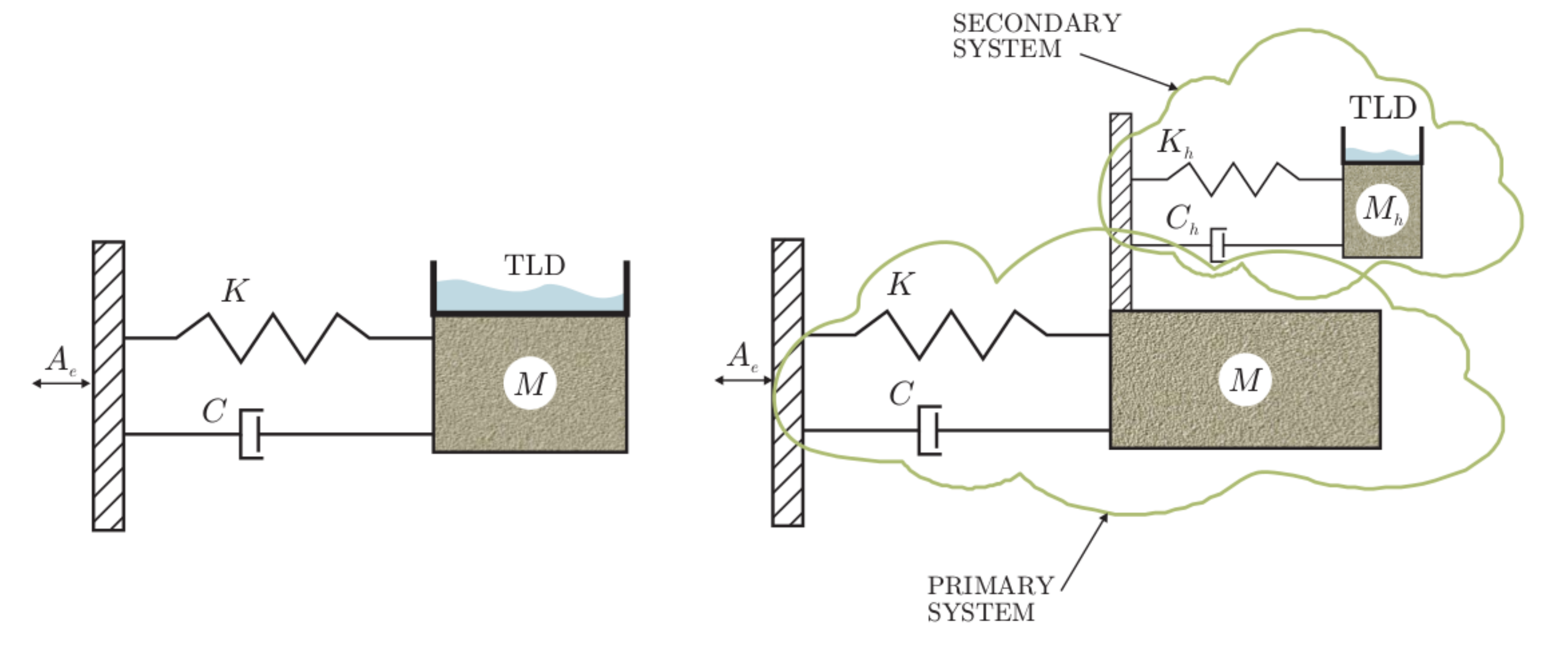}
\label{fig:TLD_HMLD}
\caption{Sketch of the two dynamical systems TLD (left) and HMLD (right) from \citep{banerji2011earthquake}. The aim is to dampen the motion of $M$ induced by an external acceleration $A_e$.}
\end{figure}

In recent decades, a certain amount of studies have been dedicated to the damping/suppression of unwanted oscillations.
This is especially true for civil infrastructures such as large buildings or bridges for which
some mechanical damping systems for structural vibration control have also been 
devised \citep{kareem_etal_1999}.
Among them, Tuned Liquid Dampers (TLD) (see left sketch of Fig. \ref{fig:TLD_HMLD}) exploit
the liquid sloshing motion in a tank in order to counteract the external forces and dissipate energy.
These dampers can be used to control a building's motion during earthquakes 
and strong winds \citep{Tamura1995,Novo_etal_2013_tldearthquake}, 
motion instabilities in spacecrafts \citep{graham1952,abramson1966} and the rolling motion in ships \citep{armenio1996b,armenio1996c,bass1998}.

The topic is receiving nowadays substantial attention in countries like Japan, where a campaign to
install dampers in existing buildings is ongoing \cite{Yamamoto_Sone_terremotos2013}. The extra weight due to the damper,
to be added to a building commonly on its top floors, often requires extremely expensive reinforcement of the building structure. For example, as reported in the media, adding mass dampers to the Shinjuku Mitsui Building
in Tokyo has been budgeted in USD 51 million. Enhancing the effectiveness of dampers, while keeping their weight low, 
becomes thus extremely important from the economic point of view.

With all these motivations, and concentrating on the TLD concept, several investigations have been performed over the years in an attempt to reproduce sloshing flows. The first studies were performed using a potential flow linear or non-linear theory, but later studies were conducted using CFD. Abundant sources can be found in the books of \citet{faltinsen_2009_cup} and \citet{ibrahim_2005}.

Modeling the energy dissipation in sloshing has always been a challenge. As an example, in order to take into account the viscosity effects and the boundary layers, some formulae for dissipation have to be added to potential flow based or shallow-water based models. In 1983, Demirbilek treated this problem of dissipation in sloshing waves both theoretically and numerically \citep{Demirbilek1983a,Demirbilek1983b,Demirbilek1983c} considering the full Navier Stokes equations. This allowed him to obtain some results regarding the influence of both Froude and Reynolds numbers on the dissipation values, but without any validation.

\citet{sun1994} performed a numerical and experimental analysis of the problem in a tank without immersed screens or structures. They identified the breaking as an important source of dissipation and determined a semi-analytical procedure to take it into account.  \citet{reed1998investigation} investigated in greater detail the effects of large amplitude sloshing on a TLD.
\citet{Marsh2011_jsv} performed experimental and numerical works regarding the analysis of dissipation mechanisms in egg-shaped sloshing absorbers, focusing on sloshing and solid boundary layer effects. From a physical point of view, the study of the dissipation induced by a free-surface flow is arduous, especially in the presence of a wave breaking flow. \citet{perlin_etal_arfm_2012_breakingwaves} presented a review and analysis on works dedicated to dissipation under wave breaking.

\citet{cooker1994water,cooker1996wave} performed elegant decay experiments with a free oscillating tank  suspended as a bifilar pendulum in the shallow-water limit, suggesting that hydraulic jump theory can provide some insight into the dissipation mechanisms.
The transfer of energy between a moving vessel and the contained fluid is studied in \citet{turner2013nonlinear}.


The calculation of resonance properties for a coupled system is often difficult in the presence of a liquid. For that, different techniques have been developed in order to model the sloshing flow and solve the coupled problem, namely \citet{yu1999non} and \citet{Tait2008} using a mass-spring subsystem, \citet{Frandsen2005} using potential flow theory and \citet{ardakani2010dynamic,ardakani2012resonance} using shallow water equations.

Recently, an alternative TLD configuration defined as a hybrid mass liquid damper (HMLD) (see \citep{banerji2011earthquake} and Fig. \ref{fig:TLD_HMLD}, right) was introduced.

The idea is to tune the mass damper $M_h$ to maximize the force counteraction between the primary and secondary structures and
to attach a sloshing damper to the mass damper in order to dissipate large amounts of energy through violent sloshing.
An optimally designed HMLD configuration is shown to be more effective as a control device than the standard TLD configuration since it maximizes
the force counteracting and dissipating effects.

In the present work, a specially devised fully coupled damper system first  described in \citet{bulian_etal_jhr09}, called hereinafter Pendulum-TLD, is analyzed. The mechanical system is essentially a non-linear driven pendulum, where the pendulum is a rectangular tank rotating around a fixed pivot.
With the tank partially filled with a liquid, energy is supplied to the whole system by a mass sliding along a linear guide fixed on the tank.
The mechanical system and the resulting sloshing flow are coupled in a very complex non-linear manner.

Part I is organized as follows:
first, the frame of reference, notation and elements of the coupled system are presented, the torques and main energy terms affecting the dynamics are identified and an analogy with TLD and HMLD systems is provided.
The dynamics of the empty tank is then described prior to developing the theoretical model representing the fluid action.
The loads of the fluid on the tank during sloshing are theoretically and numerically investigated by varying both the frequency and the amplitude of the roll motion.  Theoretical considerations are done on the scaling of the energy dissipation by the fluid. The numerical investigation is conducted using a Smoothed Particle Hydrodynamics numerical model, widely validated (see \citep{Bouscasse_etal2013,Antuono_etal_JFM2012}) in the context of violent free surface fragmentation.
The chosen model is further adapted to simulate the coupled dynamics, allowing for a non linear analysis of the coupled system behavior in the frequency domain. Conclusions are drawn and an experimental validation analysis with three different liquids is left for Part II.

\section{A pendulum Tuned Liquid Damper}\label{s:mechanicalmodel}
\begin{figure}[b!]
\centering
\includegraphics[width=0.45\textwidth]{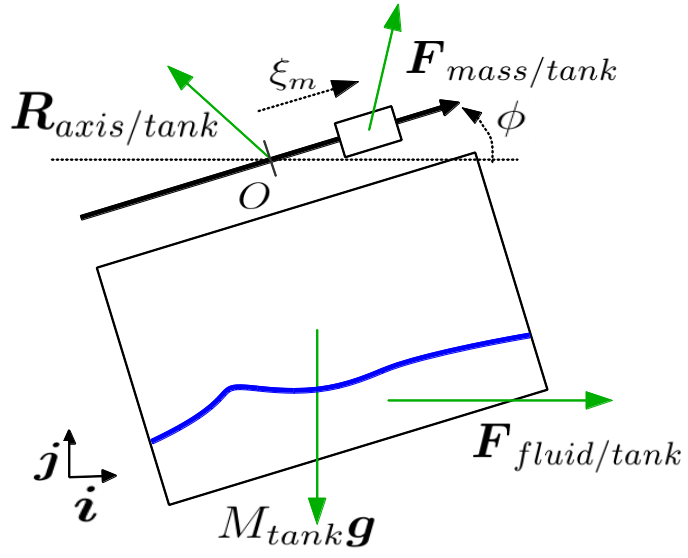}
\includegraphics[width=0.45\textwidth]{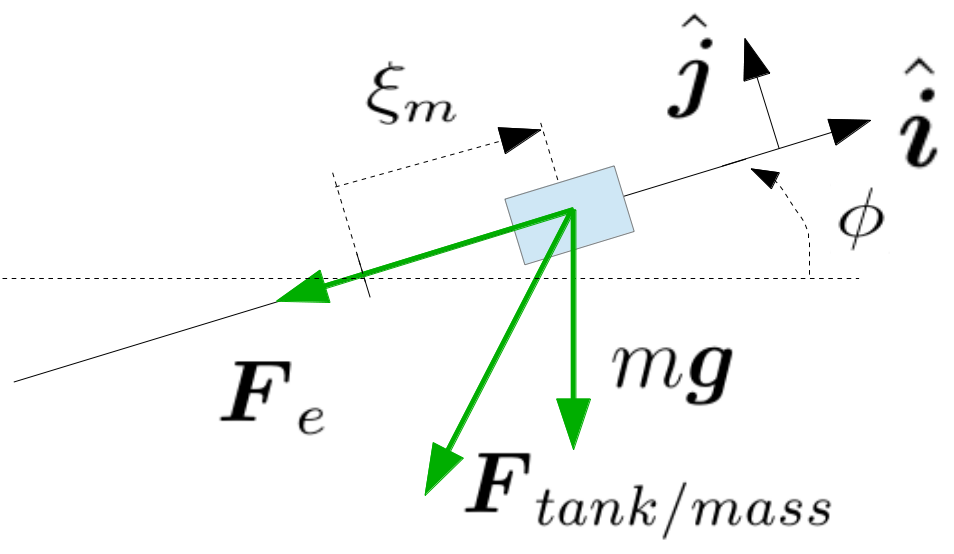}
\caption{Tank and oscillating mass (arrows are vector representation of forces). \label{fig:Tank} }
\end{figure}

The pendulum TLD (see Fig. \ref{fig:Tank}) is composed of three coupled sub-systems:
\begin{enumerate}
  \item the sliding mass,
  \item the moving parts of the sloshing rig including the empty tank but excluding the sliding mass; this sub-system will be hereafter referred to  as the tank; the energy balances will refer to this sub-system.
  \item the fluid.
\end{enumerate}

The sloshing tank is assumed to be 2D, perfectly rigid, and rotating in the vertical plane about a fixed horizontal axis passing through a fixed pivot $\vec O$. Although the findings herein are of general value, in order to conform with the experimental data of Part II,
the tank length $L$ is set equal to 0.9 m and the width $B$, normal to the plane of motion, is 0.062 m.

The length $l=0.1$ m is taken as a characteristic length of the system.
The filling height $h$ adopted and the sliding mass motion amplitude will be of this order.

The distance $H$ between the center of rotation and the tank bottom is set equal to 0.47 m.

The moment of inertia around $\vec O$ is $I_0$
and the static moment $S_G$ of the rigid system around $\vec O$ is the product
of the mass, $m_{tank}$, and the distance, $\eta_G$, between the center of gravity
of the tank and the point $\vec O$, thus $S_G = m_{tank} \eta_G$.
$I_0$ is set equal to $I_0 = 26.9$ $\mathrm{kg}.\mathrm{m}^2$ and the static moment to
$S_G = - 29.2$ $\mathrm{kg}.\mathrm{m}$ (see Part II).
The rotation center is above the center of gravity of the whole system, implying that the system is stable in the absence of external forcing.

Since the system is purely rotational, the dynamics can be described in terms of variations in angular position and through balances of angular momentum (torques) contributions. The different torques acting on the tank derive from the following four external forces: one given by the fluid $\vec F_{fluid/tank}$, another stemming from the sliding
mass $\vec F_{mass/tank}$, the third being the weight of the tank $\vec F_{static} = m_{tank} \, \vec g$ ($\vec g$ is the gravity acceleration) and the last one being the reaction of the holding structure $\vec R_{axis/tank}$ on the hinge $\vec O$ .
A scheme of these forces is provided in Fig. \ref{fig:Tank}.

The inertial frame of reference is indicated by $\left(\vec O \, \vec{i} \, \vec{j} \right)$
and the velocity $\vec u$ of a generic point $P$ on the tank is
\[ \vec{u}(P)\,=\,\dot{\phi}\,\vec{k}\, \times \vec{r} ,\]
where $\vec{k}=\vec{i} \times \vec{j}$ and is the normal vector orthogonal to the rotating plane, $\phi$
is the angular displacement, $\dot{\phi}$ is the angular velocity and $\vec r$ is the position
vector of the generic point $\vec P$ with respect to the pivot $\vec O$.

The sliding mass $m = 4.978$ kg moves along the linear guide with $\xi_m (t)$ being the coordinate along the slide. This sliding mass is forced under  a defined harmonic motion:
\begin{equation}\label{eq:sliding_mass_motion}
\xi_m(t)=A_m\,\sin(\,2\pi\,t/T),
\end{equation}
where $A_m$ is the amplitude of the oscillating mass along the linear guide, $T$ is the oscillation period.
The sliding mass motion amplitude $A_m$ is set to $0.05$, $0.10$, $0.15$ and $0.20$ m.
Since $\xi_m$ is imposed, the state of the dynamical system can be defined as
a function of the angle $\phi$ and its derivatives.

The mass moves along the axis defined by the pivot
$\vec O$ and the vector $\hat{\vec {i}}$. The non-inertial frame of reference indicated by
$\left(\vec O \, \hat{\vec{i}} \, \hat{\vec{j}} \right)$ is defined in Fig. \ref{fig:Tank}.

The forces on the sliding mass are: its weight $m \vec g$, the force given by the electric motor $\vec F_e$ and the force exerted by the tank on the mass $\vec F_{tank/mass}$. From the momentum equation in the inertial reference system
$\left(\vec O \, \vec{i} \, \vec{j} \right)$ an expression of $\vec F_{tank/mass}$ is given as a function of the
sliding mass acceleration $\vec a_m$:
\begin{equation}\label{eq:F_tank_mass}
\vec F_{tank/mass}\,=\,-m\vec g\,-\,\vec F_e\,+\,m\,\vec a_m,
\end{equation}
where $\vec a_m$ is given in $\left(\vec O \, \vec{i}\, \vec{j} \right)$ frame by:
\begin{equation}\label{eq:a_m}
\vec a_m\,=\, (\ddot{\xi_m}\,-\,\xi_m\,\dot{\phi}^2)\,         \hat{\vec i}\,+\,
              (2\dot{\xi_m}\dot{\phi}\,+\,\xi_m\,\ddot{\phi})\, \hat{\vec j}.
\end{equation}
The torque about $\vec O$ on the tank, due to the sliding mass, is:
\begin{equation}\label{eq:M_mass_tank}
\begin{array}{lll}
M_{mass/tank} &=& {\xi}_m \hat{\vec{i}} \times \vec{F}_{mass/tank} \cdot \vec{k} \,=\,\\
              &=& - m {\xi}_m  g \cos(\phi) - m(2 \xi_m \dot{\xi}_m \dot{\phi} + {\xi}_m^2 \ddot{\phi}).
\end{array}
\end{equation}
This expression comprises a term due to the weight of the sliding mass plus inertia terms
originating from the mass motion on a rotating beam.

This equation mixes together the exciting term $\xi_m$ with the roll angle $\phi$, the latter being the main output
of the dynamical system. For a sufficiently small roll angle $\phi$, a good approximation of $M_{mass/tank}$ can be given by $ - m \xi_m  g$.
The linear behavior with respect to $\xi_m$ should be dominant for $M_{mass/tank}$,
thus simplifying the analysis of the system. This hypothesis is checked with the conditions studied herein.

Following \citet{bulian_etal_jhr09} a friction torque is included in the mechanical model:
\begin{equation}\label{eq:friction}
M_{friction} \,=\, - B_{\phi} \dot{\phi} -K_{df} \sgn(\dot{\phi}),
\end{equation}
with $K_{df} = 0.54 \, \mathrm{ N.m}$ and $B_{\phi} = 0.326 \, \mathrm{ N.m}.\mathrm{(rad/s)}^{-1}$. These values have been determined in \citep{bulian_etal_jhr09} using a set of inclining and decay tests on the experimental set-up adopted in Part II of this series.

The natural frequency of the rigid system:
\begin{equation}\label{eq:natural_frq}
\omega_1^m\,=\,\sqrt{\frac{-g\,S_g}{I_0}}
\end{equation}
is equal to $\,3.263\, \mathrm{(rad/s)}$  and the corresponding period is $T_1 = 1.925\, \mathrm{s}$.
%
%
\subsection{Angular momentum and energy balances}\label{sec:Moment_Energy_Balances}
Considering the terms cited above, the angular momentum equation for the roll motion of the tank reads:
\begin{equation}\label{eq:angular_momentum}
I_0 \ddot{\phi} -  g S_g \sin(\phi) \,-\, M_{friction}\,-\,M_{fluid/tank}  \,=\, M_{mass/tank},
\end{equation}
%
where the first two terms of the left-hand side represent a classical non-linear pendulum equation
(the static moment $S_g$ has a negative value),
and the right-hand side $M_{mass/tank}$ is the forcing term of the system.

Substituting the expressions reported above, the ODE (\ref{eq:angular_momentum}) can be rewritten in an expanded form as:
\begin{equation}\label{eq:angular_momentum2}
(I_0 + m{\xi}_m^2)\ddot{\phi}  + (B_{\phi}+ 2\,m \xi_m \dot{\xi}_m) \dot{\phi}
+ K_{df} \sgn(\dot{\phi})-  g S_g \sin(\phi) + m {\xi}_m  g \cos(\phi)\,=\,\,M_{fluid/tank}\,.
\end{equation}

The torque $M_{fluid/tank}$ is given, for a Newtonian fluid, by:
\begin{equation}\label{M_fluidtank}
M_{fluid/tank}\,=\,
\displaystyle
-\,\int_{\partial \Omega_B}\, \vec{r}\times\,p\,\vec n\,dS\,+\,2\mu\,\int_{\partial \Omega_B}\, \vec{r}\times\,\mathbb{D}\,\vec n\,dS\,
\end{equation}
where the two addends represent the contribution of the pressure and the viscosity forces,
$\partial \Omega_B$ is the internal surface of the tank, $\vec n$ is the normal vector to this surface pointing away from the fluid, $\mathbb{D}$ is the fluid velocity strain rate tensor and $\mu$ the dynamic viscosity of the fluid.
For an ideal fluid in absence of breaking, it is generally possible to find an analytical or semi-analytical expression
for eq. (\ref{M_fluidtank}) (see \emph{i.e.} section \ref{sec:Verhagen}), while for the more general case, $M_{fluid/tank}$ can
only be found through a numerical solver (see  \emph{i.e.} subsection \ref{SPH}).
A proper choice of the fluid characteristics and filling height should allow
for a fluid response $M_{fluid/tank}$ that can suppress unwanted tank oscillations excited by external forces.

Equation \ref{eq:angular_momentum} can be multiplied by the angular velocity $\dot{\phi}$ and integrated over an oscillation period
to obtain the following energy balance:
\begin{equation}\label{eq:energy_balance_1}
[E_{tank}^{mech}]_t^{t+T}\,-\,\Delta\,E_{friction}\,-\,\Delta\,E_{fluid/tank}\,\,=\,\Delta\,E_{mass/tank}
\end{equation}
in which:
\begin{enumerate}
  \item $\Delta\,E_{mass/tank}$ accounts for the energy transfer between the sliding mass and the tank in one cycle; it is defined as:
\begin{equation}\label{eq:deltaEmasstank}
\Delta\,E_{mass/tank}\,=\,\int_{t}^{t+T}\,M_{mass/tank}(s)\,\dot{\phi}\,ds.
\end{equation}
The sliding mass is the driving element of the system and  $\Delta\,E_{mass/tank}$ is therefore expected to be positive.
In reality, if the damping phenomena are not energetic enough, there may be cycles for which there is a net transfer of energy from the tank to the moving mass,
as shown in the next sections.
  \item $[E_{tank}^{mech}]_t^{t+T}$ is the variation of the mechanical energy of the tank during one oscillation cycle; it is defined as:
\begin{equation}
[E_{tank}^{mech}]_t^{t+T}\,:=\,\int_{t}^{t+T}\,\left(I_0 \ddot{\phi} -  g S_g \sin(\phi)\right)\,\dot{\phi}\,ds,
\end{equation}
  \item $\Delta\,E_{friction}$, always negative, is the energy variation of the tank due to the mechanical
  friction for one cycle; it is defined as:
\begin{equation}
\Delta\,E_{friction}\,:=\,\int_{t}^{t+T}\,M_{friction}(s)\,\dot{\phi}\,ds,
\end{equation}
  \item $\Delta\,E_{fluid/tank}$ is the energy transfer between the fluid and the tank during one cycle; it is defined as:
\begin{equation}\label{eq:E_fluid_tank}
\Delta\,E_{fluid/tank}\,:=\,\int_{t}^{t+T}\,M_{fluid/tank}(s)\,\dot{\phi}\,ds.
\end{equation}
This term is linked to the sloshing phenomena induced by the tank motion.
In order to dampen such a motion, $\Delta\,E_{fluid/tank}$ should be negative, that is, in one period of oscillation the tank exerts a positive work on the fluid and not vice versa; this issue will further be discussed in detail in the rest of the paper.
\end{enumerate}
%
The energy variation $\Delta\,E_{fluid/tank}$ is characterized by two components:
\begin{enumerate}
  \item $[E_{fluid}^{mech}]_t^{t+T}$ is the mechanical energy balance of the fluid in one oscillation cycle.
  \item $\Delta\,E_{fluid}^{dissipation}$ is the energy dissipated by the fluid in one cycle and is always negative (see \emph{e.g.} \citep{aris1989vectors}).
\end{enumerate}
The energy $\Delta\,E_{fluid}^{dissipation}$ involves different phenomena:
(i) the fluid friction on the tank walls,
(ii) water impacts against the vertical walls,
(iii) breaking waves.
The magnitudes of these different components depend on the nature of the fluid. For example, when using water, breaking waves are expected to be the main source of fluid dissipation.

The energy balance for the fluid hence reads:
\begin{equation}\label{eq:P_fluid_dissipation}
\Delta\,E_{fluid/tank}\,=\,-[E_{fluid}^{mech}]_t^{t+T}\,+\,\Delta\,E_{fluid}^{dissipation}
\end{equation}

and therefore equation (\ref{eq:energy_balance_1}) becomes:
\begin{equation}\label{eq:Energy_balance}
 [E_{tank}^{mech}]_t^{t+T}\,+\,[E_{fluid}^{mech}]_t^{t+T}\,-\,\Delta\,E_{friction}\,-\,\Delta\,E_{fluid}^{dissipation}\,=\,\Delta\,E_{mass/tank}
\end{equation}
The work done by the sliding mass, when positive, increases the mechanical energy of the tank and fluid,
but is partially dissipated by the mechanical friction and fluid dissipation mechanism.

\begin{figure}[hb] 
\centering
\vspace{0.5cm}
\includegraphics[width=0.60\textwidth]{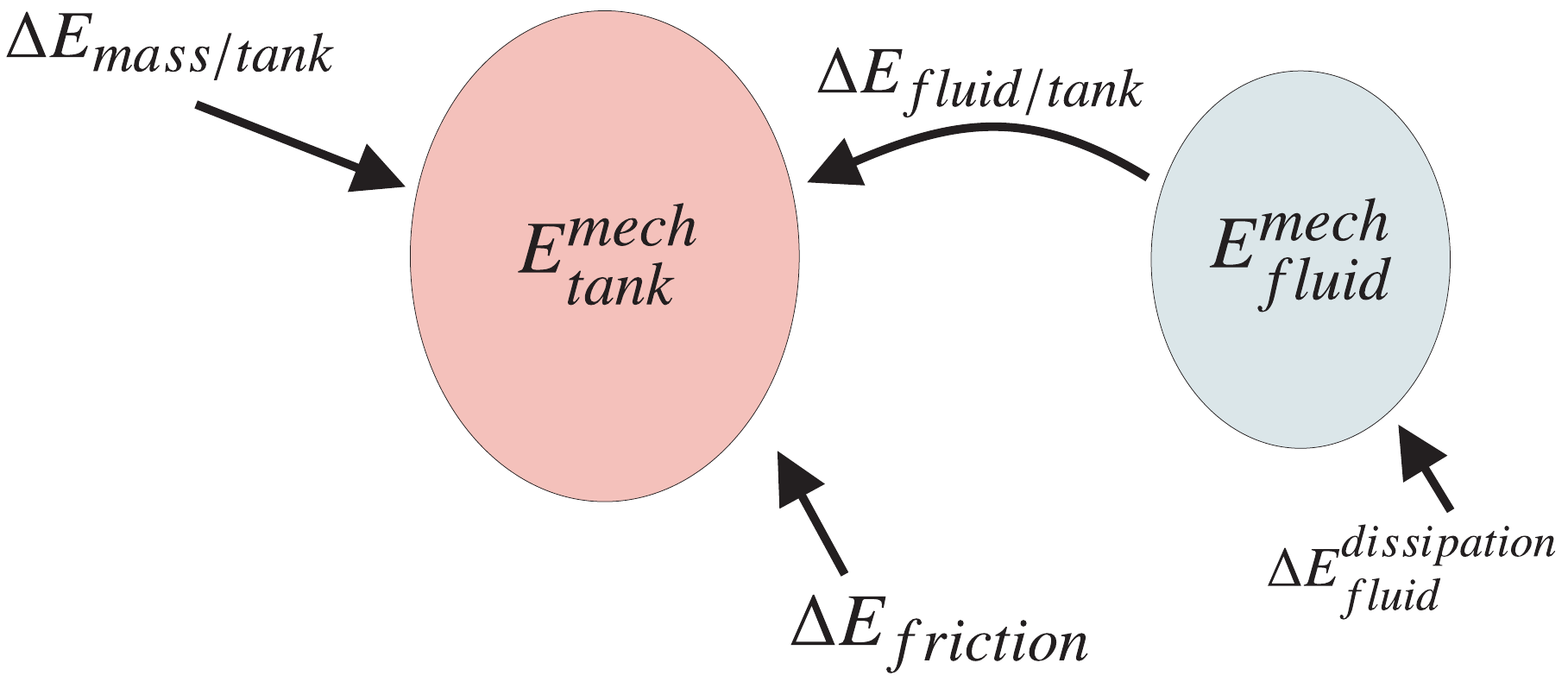}
\caption{Energy balance between the sliding mass, the tank and the fluid; the direction of the arrow corresponds to positive contributions.}
\label{fig:DeltaEs}
\end{figure}

All the energy contributions are represented in Fig. \ref{fig:DeltaEs}, where the direction of the arrow indicates the positive sign contribution.
%
%
\subsection{Analogies between the present system, a TLD and a HMLD}\label{sec:xxxx}
\subsubsection{TLD}
The amplitude of the roll angle $\phi$ is the main indicator
of the performance of a TLD system with angular motion.
For a given excitation $A_m$, the lower the $\phi$ the  more effective the TLD is considered to be.
An analogy can be established between the present system and an angular motion
TLD. Looking at the left panel of Fig. \ref{fig:TLD_HMLD}, mass $M$ can be thought
of as tank for the system described herein, the TLD being the fluid, $K$ being the restoring term of the moment equation,
$C$ the friction term, and $A_e$ the moment due to the moving mass.

However, this analogy falls short because, as will later be seen,
if the excitation is above a certain threshold,
the roll angle may not be reduced even if the system is dissipating a large amount of energy.
This property suggests the idea of looking at the system as a hybrid mass liquid damper (HMLD).
\subsubsection{HMLD}
\label{sss:hmld}
A large proportion of the analysis for the present system is in the energy transfer between the moving
mass and the tank, $\Delta E_{mass/tank}$. Looking at the right panel of Fig. \ref{fig:TLD_HMLD},
the present system can be seen as the secondary system (with mass $M_h$) and may experience a larger amplitude motion than
permitted if attached to the main structure. Under this
large motion scenario, high levels of energy transfer may be induced from the primary damper $M$
through the term $\Delta E_{mass/tank}$, finally being dissipated on the secondary system with the
large angular motion sloshing flows.
%
\subsection{Definition of envelopes and phase lags functions}\label{sec:Preliminary_Consideration}
In this subsection, useful quantities are defined in order to properly analyze the present system. These quantities are not used to obtain analytical solutions but only to extract important information from the numerical simulations and experimental results.
As an example, the solution of equation (\ref{eq:angular_momentum2}), $\phi(t)$, can be approximated as:
\begin{equation}\label{eq:phi_form1}
\phi(t)\,=\,\Phi_{env}(t)\,\sum_{n=1}^{\infty} \sin\,[n\,\omega\,t\,+\,\delta_n(t)\,] \, ,
\end{equation}
%
provided that the envelope function $\Phi_{env}(t)$ and phase shift functions $\delta_n(t)$ each have
slow dynamics with respect to the period $T=2 \pi /\omega$. 
As will be shown later, the first harmonic component is largely dominant in the roll motion.
Therefore, $\phi(t)$ can be described with a good approximation by:
\begin{equation}\label{eq:phi_form}
\phi(t)\,\approx  \,\Phi_{env}(t)\,\sin\,[\,\omega\,t\,+\,\delta(t)\,] \, .
\end{equation}
%
Due to their slow dynamics, the envelope function $\Phi_{env}(t)$ can be approximated as:
\begin{equation}\label{eq:phi_eff}
\Phi(t)\, = \,\frac{\pi}{2\,T}\int_{t}^{t+T}\,|\phi(s)|\,ds\,,
\end{equation}
and the shift function $\delta(t)$ can be approximately evaluated looking at the maximum values of $\phi(t)$ and $\xi_m(t)$ in a moving $T-$time window, and measuring the relative time shift in order to evaluate the phase lag.

Due to the existence of dissipative terms, equation (\ref{eq:angular_momentum2}) may admit a a ``time-periodic solution'' and equation (\ref{eq:phi_form}) becomes:
\begin{equation}\label{eq:phi_steadystate_0}
\phi(t)\, =\,\Phi \sin [\omega t \,+\,\delta]
\end{equation}
This is true for a large number of conditions, however,  it is known \citep{Bouscasse_etal2013} that shallow water sloshing can lead to subharmonics, in particular with low amplitude oscillations.
Regarding the torque $M_{fluid/tank}$, at time-periodic state, equation (\ref{eq:angular_momentum2}) shows that even considering the approximation (\ref{eq:phi_steadystate_0}) for the roll angle,
the non-linear terms induce non-negligible effects on the time behavior and the torque exerted by the fluid on the tank needs to be expressed as:
\begin{equation}\label{Mfluid_verhagen1}
M_{fluid/tank} =\sum_{n=1}^{\infty} M_n \sin [\,n (\omega t \,+\,\delta)\,+\,\Psi_n\,]
\end{equation}
where $\Psi_n$ is the phase lag between the torque $n-$harmonic component and the roll angle $\phi(t)$ and
in which, the first harmonic component is expected to play a lead role in the time-periodic state balance.

Prior to the time-periodic state, and similarly to what is done with $\delta(t)$,
it is possible to extract a function $\Psi(t)$ from
the time histories of $M_{fluid/tank}(t)$, looking for the local maximum.
Indeed, this phase lag function evolves with slow time dynamics with respect
to the period $T$.
%

\begin{figure}[ht!]
\centering
\includegraphics[width=0.72\textwidth]{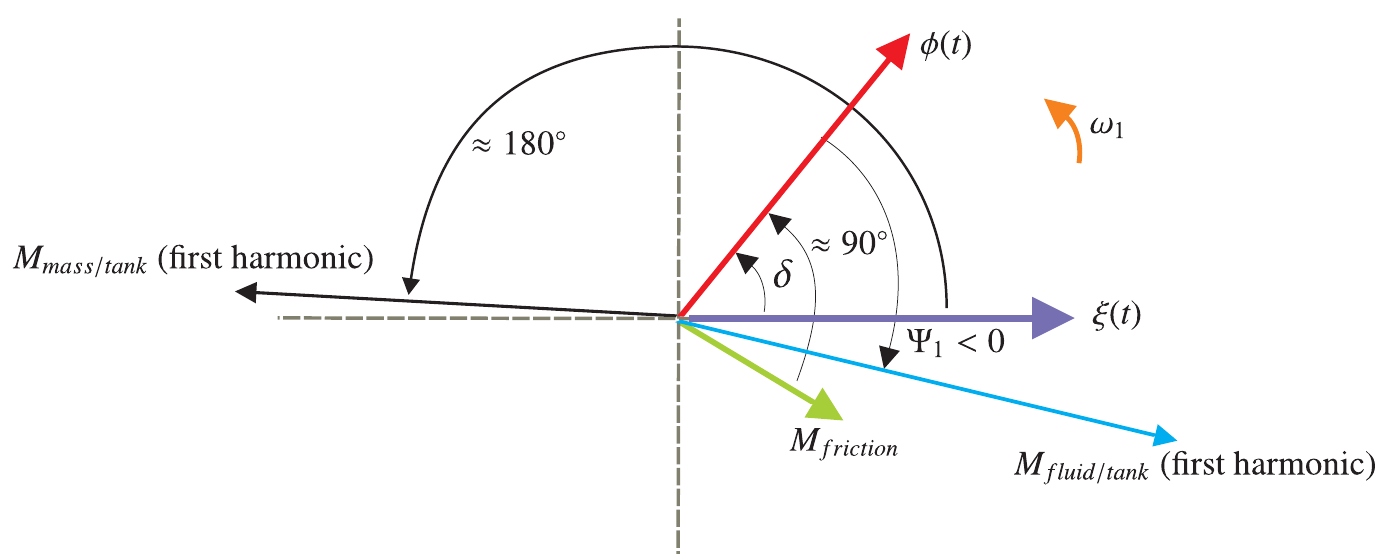}
\caption{
Complex plane: main torques and motions involved in the analysis.
}
\label{fig:momentosyangulos}
\end{figure}

Summarizing, in the time-periodic state, it is possible to define phasors on a complex plane using the modulus and phases of the different quantities. The origin for the phases is given by the sliding-mass motion.
In order to help in assimilating the notation and in identifying the main actors of the dynamics under study,
a typical configuration at time-periodic state is sketched in Fig. \ref{fig:momentosyangulos}.
%

The following observations can be made:
\begin{enumerate}
  \item Typically, the tank motion, $\phi(t)$, is lagged with respect to the sliding mass motion, $\xi(t)$, with an angle, $\delta$, smaller than $90^\circ$.
  \item The torque created by the sliding mass, $M_{mass/tank}$, is lagged approximately $180^\circ$ with respect to the sliding mass motion $\xi(t)$.
  \item The torque due to the friction term is advanced approximately $90^\circ$ with respect to the tank motion $\phi(t)$.
  \item The optimum condition in order to damp the tank motion takes place when the torque $M_{fluid/tank}$ acts in counter-phase with respect to the torque $M_{mass/tank}$.
  Fulfillment of this condition is discussed in section \ref{sec:Verhagen}.
\end{enumerate}
%
\section{Dynamics of the system with the empty tank}\label{empty_tank}
\subsection{General} \label{empty_tank:gen}
In this section the system is studied without fluid, focusing on the dependencies of the moving mass amplitude $A_m$ and the effect of friction on the dynamics.

Considering an empty tank and null friction term, equation (\ref{eq:angular_momentum2}) can be reduced to:
%
\begin{equation}\label{eq:non_linear_pendulum}
\left[1\,+\,\frac{m A_m^2}{I_0}\sin^2(\omega t) \right]\,\ddot{\phi}  \,+\,
\frac{m A_m^2}{I_0} \omega \sin(2\omega t)\, \dot{\phi} + {\omega_{1m}}^2 \sin(\phi) \,+\,\frac{mgA_m}{I_0}\sin(\omega t) \cos(\phi)\,=\,0
\end{equation}

As previously mentioned, equation (\ref{eq:non_linear_pendulum}) has practically the same behavior as a
driven non-linear pendulum. The dynamics of the empty tank condition is explored numerically looking at this ODE (\ref{eq:non_linear_pendulum}). The accuracy of this model of the ``empty-tank'' behavior was demonstrated in \citep{bulian_etal_jhr09}.

\begin{figure}[ht!]
\centering
\includegraphics[width=0.8\textwidth]{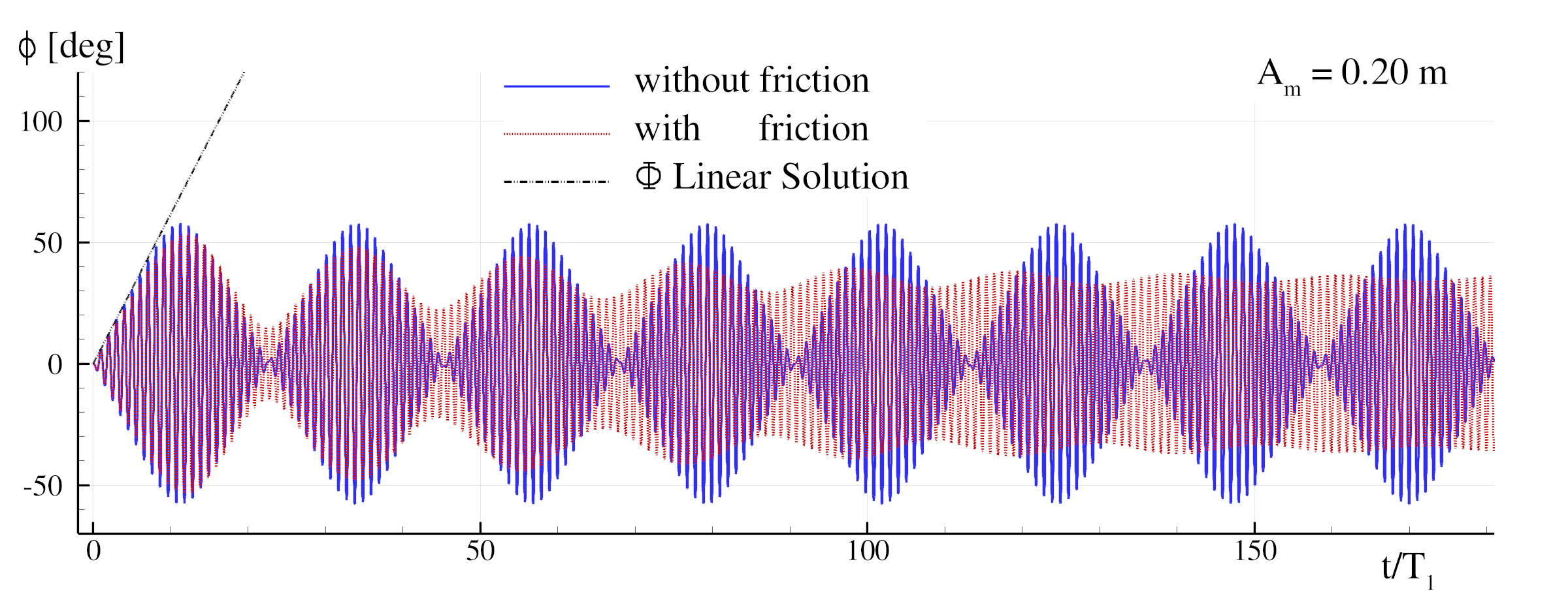}
\caption{Empty tank model: Roll angle $\phi$ plotted as a function of time using an excitation amplitude $A_m = 0.20$ m, $\omega=\omega_1^m$.
Solid line: without friction, dashed line: with the friction model.
} \label{fig:empty_tank_phi}
\end{figure}

In Fig. \ref{fig:empty_tank_phi}, the solid line refers to the solution of equation (\ref{eq:non_linear_pendulum})
using the largest amplitude of excitation $A_m = 0.20$ m for the sliding mass and $\omega_1^m$ as excitation frequency.
The solution shows the classical beating characteristic of a driven non-linear pendulum (see {\emph e.g.} \cite{butikov2008extraordinary}).

The energy exchanged between the tank and the sliding mass periodically changes in sign during the beating periods.

In the initial part of the time histories plotted in Fig. \ref{fig:empty_tank_phi}, $\phi(t)$ essentially follows
the linear resonant solution:
\begin{equation}\label{linear_resonance}
\phi_{Lin}(t)\,=\,\frac{m\,g\,A_m}{I_0}\frac{t\,\cos(\omega_1^{m} \,t)}{2 \omega_1^m}\,
\end{equation}
in which the amplitude $\Phi(t)$ grows linearly with time and $\phi_{Lin}(t)$
is in quadrature (i.e. 90 degrees out of phase) with the sliding mass $\xi_m(t)$  .

When considering the friction of the system, the solution (see dashed line in Fig. \ref{fig:empty_tank_phi})
shows that $\phi$ reaches a time-periodic state after a long transient.
A similar behavior is expected when the fluid is in the tank. Indeed, the dissipation mechanisms of the fluid added to the friction mechanism should generate a time-periodic state in a shorter time range.

\begin{figure}[ht!]
\centering
\includegraphics[width=0.8\textwidth]{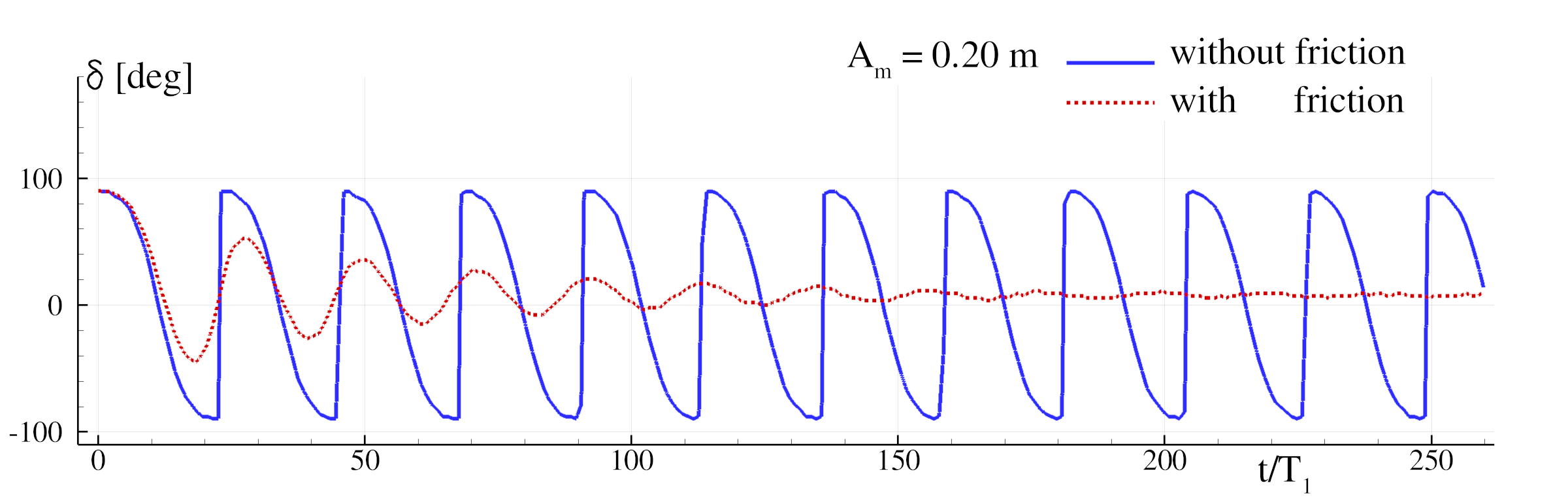}
\caption{Empty tank model: shift function $\delta$ plotted as a function of time using an excitation amplitude $A_m = 0.20$ m,$\omega=\omega_1^m$.
Solid line: without friction, dashed line: with the friction model.
} \label{fig:empty_tank_delta}
\end{figure}

Figure \ref{fig:empty_tank_delta} shows the shift function $\delta(t)$ plotted as a function of time in the empty tank condition ($A_m = 0.20$ m
with and without friction terms). Without friction, the phase lag oscillates periodically between $90^\circ$ and $-90^\circ$ (solid line). With friction, the oscillations of $\delta(t)$ decrease in time towards a constant value ($8^\circ$).

\begin{figure}[ht!]
\centering
\includegraphics[width=0.8\textwidth]{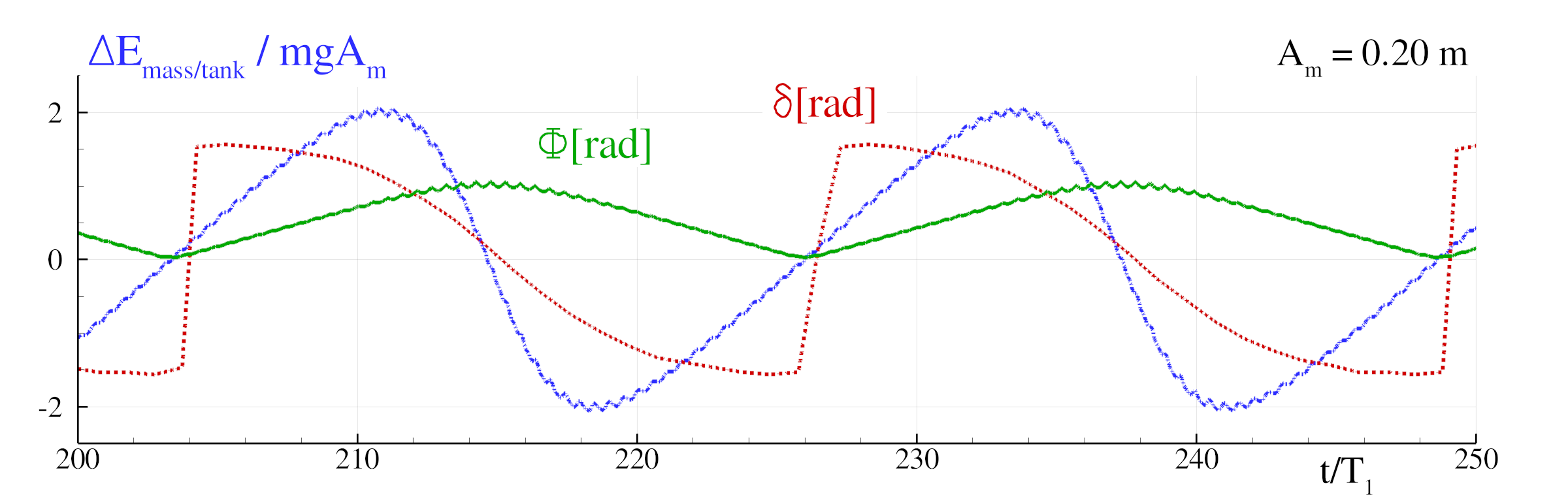}
\includegraphics[width=0.8\textwidth]{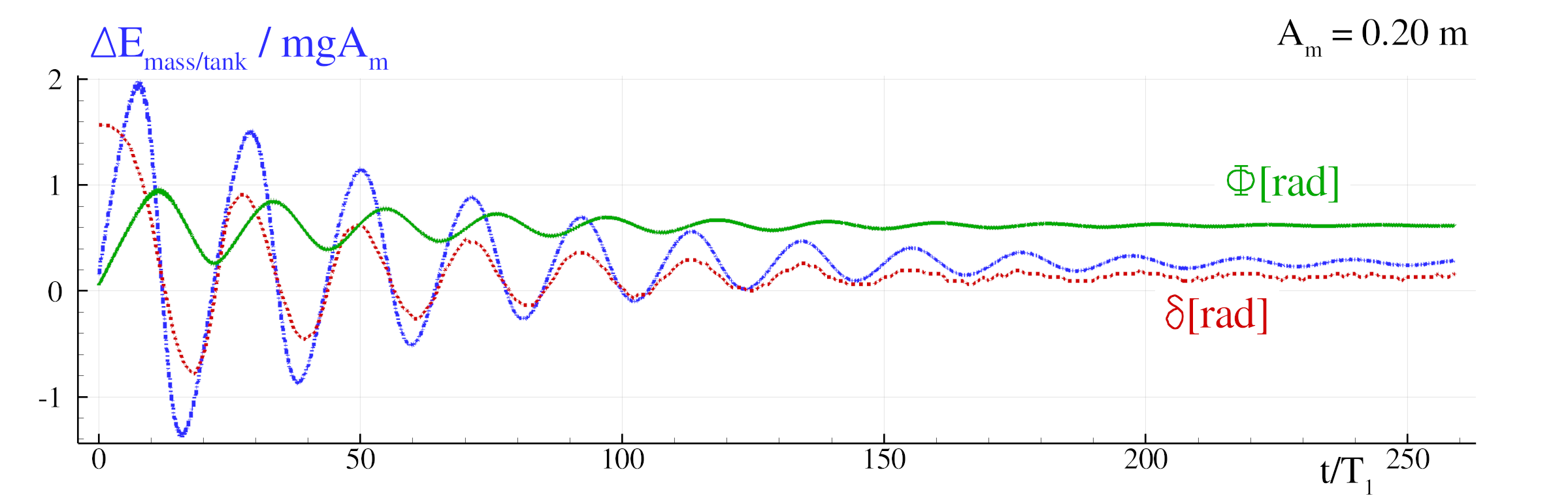}
\caption{Empty tank model: $\Delta E_{mass/tank}$ plotted as a function of time using an excitation amplitude $A_m = 0.20$ m, $\omega=\omega_1^m$.
Top: without friction terms, bottom: with friction terms.}
\label{fig:empty_tank_Energy}
\end{figure}

For positive values of $\delta$, the mass is transmitting energy to the tank whilst for negative $\delta$,
the tank gives back some energy to the sliding mass ({\em i.e.} $\Delta E_{mass/tank}<0$).
The behavior of $\Delta E_{mass/tank}$, $\Phi$ and $\delta$ in time is better depicted by the plots in Fig. \ref{fig:empty_tank_Energy}.


\begin{figure}[ht!] 
\centering
\vskip 0.2cm
\includegraphics[width=0.99\textwidth]{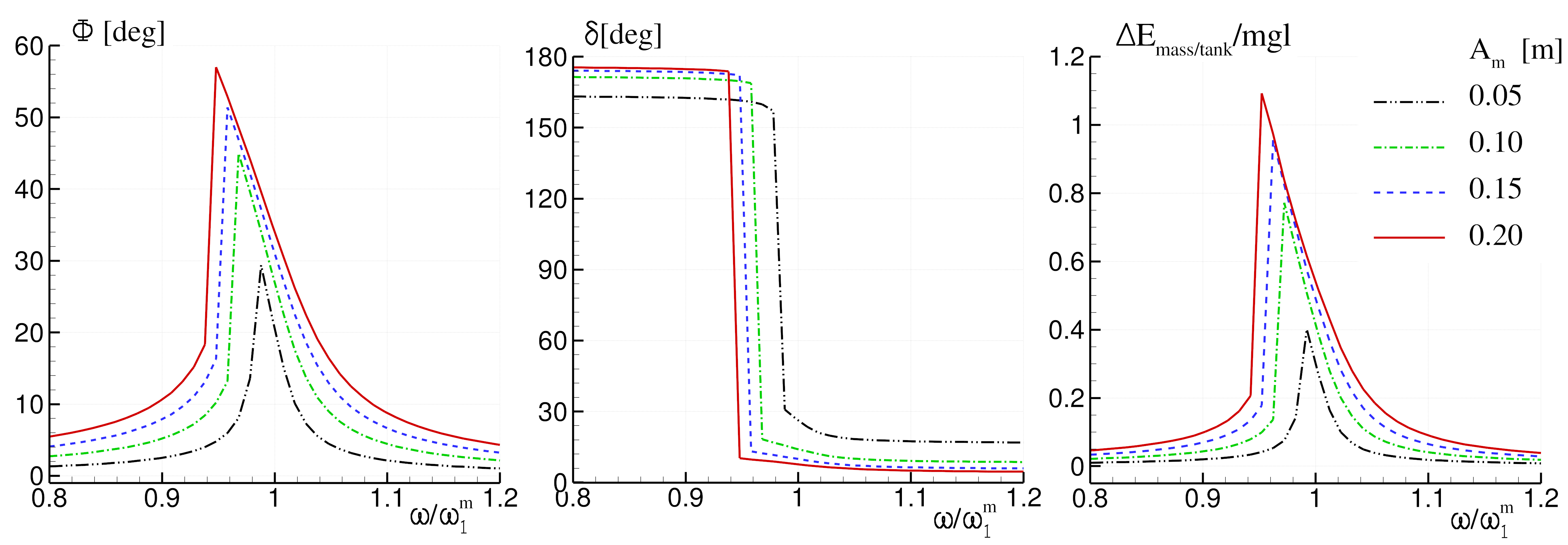}
\caption{Empty tank model with friction terms: roll angle $\Phi$, phase lag $\delta$ and $\Delta E_{mass/tank}$
reached at a time-periodic state for different excitation frequencies.}
\label{fig:Operator_Vacio}
\end{figure}

Fig. \ref{fig:Operator_Vacio} shows the frequency behavior of $[\Phi, \delta, \Delta E_{mass/tank}]$ at time-periodic state. This plot highlights the non-linearity of the mechanical system with the typical bifurcation phenomenon on $\Phi$ when varying the frequency $\omega$ (see {\emph e.g.} \cite{butikov2008extraordinary}).
Also, when increasing $A_m$, the frequency at which the maximum $\Phi$ appears (\emph{i.e.} at which $\delta=90^\circ$) moderately decreases and is lower than $\omega_1^m$ . This ``soft spring" behavior is well documented in the literature.

\begin{table}[ht!]
\begin{center}
    \begin{tabular}{|l||c|c|c|c|}
        \hline
        $A_m$ [m]                   & 0.05   & 0.10  & 0.15   & 0.20    \\ \hline \hline
        $\Phi$ [degree]             & 20     & 27    & 31     & 34      \\ \hline
        $\delta$ [degree]           & 26     & 14    & 10     & 8       \\ \hline
        $\Delta E_{mass/tank}/mgl$  & 0.30   & 0.42  & 0.50   & 0.54    \\ \hline
    \end{tabular}
\caption{Empty tank model: values of the main quantities reached at time-periodic state for the excitation amplitudes: $A_m$= 0.05, 0.10, 0.15 and 0.20 m and using the excitation frequency $\omega\,=\,\omega_1^m$.\label{tab:empty}}
\end{center}
\end{table}

Table \ref{tab:empty} reports the value of $[\,\Phi\,,\delta\,,\Delta E_{mass/tank}\,]$
reached at a time-periodic state using $\omega\,=\,\omega_1^m$. Those values are to be used
as reference data for Part II where the tank is filled with a liquid.
\subsection{Torque exerted by the sliding mass on the empty tank}
\begin{figure}[ht]
\centering
\vskip 0.2cm
\includegraphics[width=0.90\textwidth]{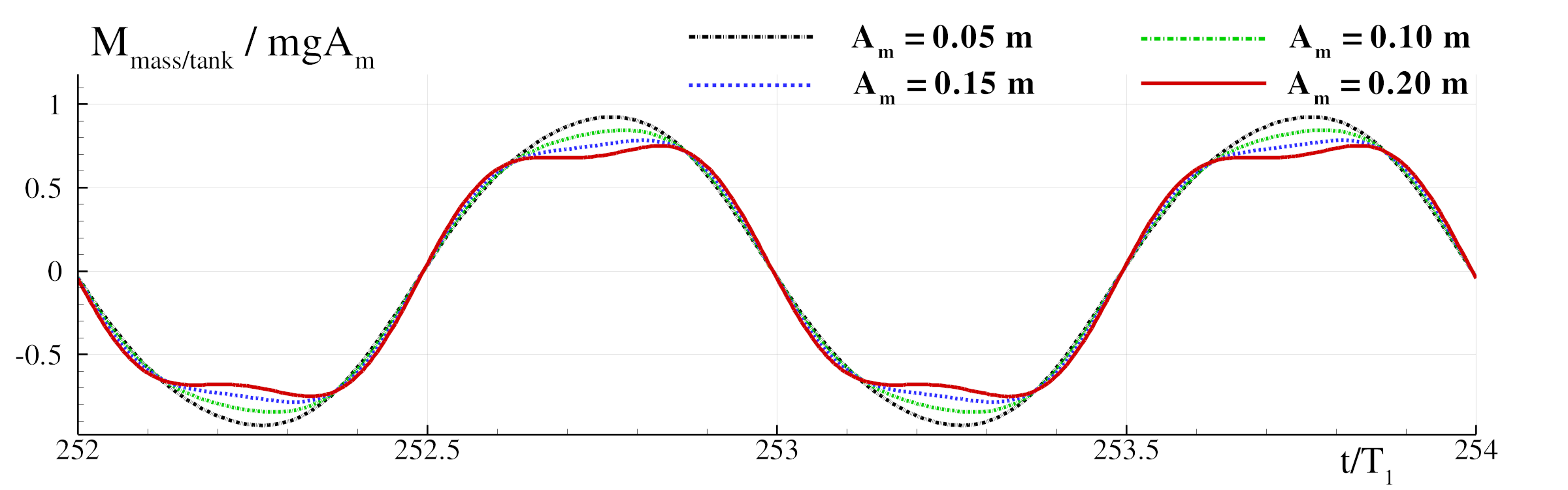}
\caption{Empty tank model: $M_{mass/tank}$ is plotted as a function of time during a time-periodic state where $\omega=\omega_1^m$
for four excitation amplitudes.}
\label{fig:EmptyTank_Mmasstank}
\end{figure}

Figure \ref{fig:EmptyTank_Mmasstank} shows $M_{mass/tank}/{mgA_m}$ as a function of time
for $\omega=\omega_1^m$ at a time-periodic state (when the time-periodic state is met).
This torque is a non-linear function of $\xi(t)$, $\phi(t)$ and their derivatives
(see eq. \ref{eq:M_mass_tank}).

The mentioned figure highlights the effect of increasing $A_m$.
For the lowest $A_m = 0.05$ m, the torque is almost sinusoidal.
When increasing the excitation amplitude it remains in phase
with the sliding mass motion $\xi(t)$. It is also noticeable from the figure that a
saturation effect takes place on the upper/lower parts of the signal when
$A_m$ is increased.
%
%
\section{Theoretical and Numerical predictions of the torque exerted by the fluid and the associated dissipation}\label{sec:Verhagen}
\subsection{General}
The resonance characteristics of a sloshing tank subjected to swaying and rolling have been thoroughly investigated over the years.
The torque $M_{fluid/tank}$ (see equation \ref{M_fluidtank}) depends on the value of the pressure and velocity fields. It is not possible to find a general formulation in closed form for the Navier Stokes equations, especially when free surface breaking occurs.

\begin{figure}[ht!]
\centering
\includegraphics[width=0.99\textwidth]{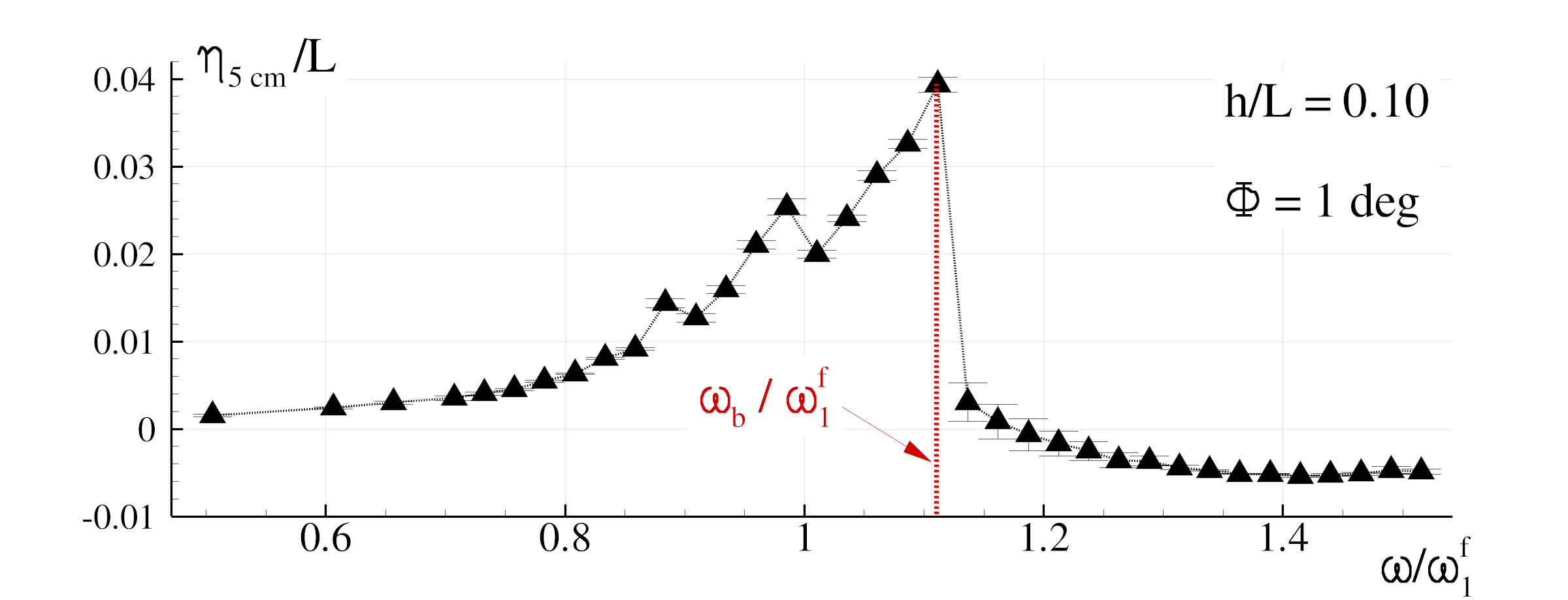}
\caption{
Maximum wave elevation measured at 0.05 m from the
left vertical wall obtained during the time-periodic
state varying the excitation frequency $\omega$.
The roll angle amplitude is set to $\Phi=1^\circ$.
}
\label{fig:Sloshing_Operator}
\end{figure}

Figure \ref{fig:Sloshing_Operator} shows a typical frequency behavior of the wave amplitude during periodic sloshing
in a rectangular tank for shallow water conditions.
Increasing the excitation frequency raises the wave elevation until a frequency $\omega_b$ where a bifurcation is observed.
For frequencies $\omega>\omega_b$ the wave elevation is drastically reduced.
In shallow water condition $\omega_b$ is always larger than $\omega_1^f$ (see \cite{faltinsen2000} and \cite{Bouscasse_etal2013}),
where $\omega_1^f$ is the first natural sloshing frequency:
\begin{equation}\label{First_Sloshing_resonance}
\omega_1^f\,= \sqrt{\,g\, \pi/L \tanh(\pi h/L)}.
\end{equation}
Therefore, the sloshing flow intensity has a ``hard spring" type amplitude response,
the opposite of the ``soft spring" behavior of $\Phi$ discussed in section \ref{empty_tank}
for the empty tank condition. In Fig. \ref{fig:Sloshing_Operator}, small peaks are visible
on the wave amplitude measurements when $\omega<\omega_b$.
Those are related to secondary resonance effects, which are typical phenomena in shallow water sloshing dynamics
(for more details see \cite{chester1968resonantI}, \cite{chester1968resonantII}).

%
\subsection{Torque from Verhagen and Van Wijngaarden analysis}
\label{ss:verhagenmfluid}
In the pioneer work of \citet{verhagen1965}, the non-linear inviscid problem is solved for a shallow water regime using hydraulic jump solutions on a tank forced in roll motion with a harmonic time history:
\[
\phi(t)\,=\,\Phi\,\sin(\omega\,t\,+\,\delta),
\]
with a constant $\Phi$ and an arbitrary phase $\delta$.

According to \citet{verhagen1965}, hydraulic jumps travelling back and forth between the walls of the tank
exist in the following range of excitation frequencies:
\begin{equation}\label{verhagen}
\big(\omega\,-\,\omega_1^f\big)^2\,<\,\frac{24\,g\,\Phi}{L}.
\end{equation}
$M_{fluid/tank}$ can be expressed with Fourier series:
\begin{equation}\label{Mfluid_verhagen}
M_{fluid/tank} =\rho g \Bigg(\frac{L}{2}\Bigg)^3 \,B\,\sum_{n=1}^{\infty} \tilde{M}_n \sin [\,n (\omega t \,+\,\delta)\,+\,\Psi_n\,],
\end{equation}
with $\Psi_1$ being the phase lag between the first harmonic component of the torque and
the roll angle $\phi(t)$ (see Fig. \ref{fig:momentosyangulos}).

In the vicinity of the resonance frequency $\omega_1^f$, the first harmonic component in equation \ref{Mfluid_verhagen} is given by:
\begin{equation}\label{eq:M_Verhagen}
\left\{
\begin{array}{lll}
\ds
\tilde{M}_1 &=& \ds
         \Bigg( \frac{2}{3}      \Bigg)^{\frac{3}{2}}\,
         \Bigg( \frac{4}{\pi}    \Bigg)^4\,
         \Bigg( \frac{\Phi\,h}{L}\Bigg)^\frac{1}{2}\,
         \Bigg[\,1\,-\,\frac{L(\omega\,-\,\omega_1^f)^2}{32\,g\,\Phi}\Bigg]
         \\[0.50cm]
\Psi_1 &=& \ds\,-\frac{\pi}{2}
           \,-\,2\arcsin \Bigg[ \frac{L(\omega\,-\,\omega_1^f)^2}{24\,g\,\Phi}
                         \Bigg]^\frac{1}{2}
           \,+\,\arcsin  \Bigg[ \frac{L(\omega\,-\,\omega_1^f)^2}{96\,g\,\Phi\,-\,3L\,(\omega\,-\,\omega_1^f)^2}
                         \Bigg]^\frac{1}{2}\,.
         \\[0.3cm]
\end{array}
\right.
\end{equation}
%
Therefore, in the \citet{verhagen1965} analysis, the torque magnitude is proportional to $\sqrt{\Phi}$
and its maximum value is achieved for $\omega=\omega_1^f$, i.e. when the system is forced with the fluid
resonance frequency.

From equation (\ref{eq:M_Verhagen}) and for values of $\omega$ in accordance with equation (\ref{verhagen}), $\Psi_1$
decreases from $0$ to $-180^\circ$. Specifically for $\omega=\omega_1^f$,  $\Psi_1=-90^\circ$, which
implies that the first harmonic of $M_{fluid/tank}$ is in quadrature with the tank motion (see equation \ref{Mfluid_verhagen}).
%
\subsection{Theoretical fluid dissipation}\label{theo_fluidDiss}
A theoretical approximation of the fluid energy dissipation has now been developed similar to
\citep{verhagen1965}, where hydraulic jump solutions for an inviscid flow are used.
Following \citep{Stoker_1957}, the energy loss dissipated across a hydraulic jump between water heights
$h_0$ and $h_1$ moving with velocity $u$ over a wave period is:
\begin{equation}
\Delta E_{fluid}^{dissipation}   \approx  - B \rho g h_0 u \frac{(h_1-h_0)^3}{4 h_0 h_1} T.
\end{equation}
Over one period the total distance that the wave must propagate down the length of the tank and back again, is $uT=2L$. Therefore, the above expression becomes:
\begin{equation}
\Delta E_{fluid}^{dissipation}  \approx  - 2 B  L\rho g h_0 \frac{(h_1-h_0)^3}{4 h_0 h_1}.
\end{equation}
All of the energy given to the fluid comes from the motion of the tank's walls. For an inviscid fluid this can be approximated to the work done by pistons acting against the net difference between the (approximately hydrostatic) pressure distributions at the two end walls.
For an elementary angle rotation $d \phi$ the work is:
\begin{equation}
dW_p =  - \frac{\rho g B}{2} (h_1^2-h_0^2) \sqrt{H^2+(L/2)^2} d\phi
\end{equation}
Integrating over an oscillation period and since this value should be equal to $\Delta E_{fluid}^{dissipation}$,
on one hand $(h_1-h_0)$ is obtained combining above expressions, on the other hand $h_1+h_0 \approx 2h$,
and subsequently the following estimation for $\Delta E_{fluid}^{dissipation}$ can be given:
%
\begin{equation}\label{Energy_Verhagen_Ben}
\begin{array}{lll}
\ds
\Delta E_{fluid}^{dissipation}&=& \ds -[(2H/L)^2\,+\,1]^{3/4}\,(4\rho\,g\,\,B\,L\,h^2)\,\Phi^{\frac{3}{2}}\,=\,
\\[0.20cm]
&=& \ds
-[(2H/L)^2\,+\,1]^{3/4}\,(4\,m_{liquid}\,g\,h\,)\,\Phi^{\frac{3}{2}}
\end{array}
\end{equation}
where $m_{liquid}$ is the mass of the sloshing liquid contained in the tank.
Since for the present system $2H/L\simeq\,1$, the above expression reduces to:
\begin{equation}\label{Energy_Verhagen}
\frac{\Delta E_{fluid}^{dissipation}}{(4\,m_{liquid}\,g\,h\,\Phi^{\frac{3}{2}})}\,=\,-  2^{3/4}\approx -1.68
\end{equation}
%
This dissipation rate is constant in time, which is generally not the case for a real sloshing flow where
a breaking wave front develops only on limited time ranges and is not present during the whole oscillation cycle.
This is the reason why equation (\ref{Energy_Verhagen}) tends to over-predict the fluid dissipation as shown in the next subsection.

This non-dimensional coefficient linked to the energy dissipated by the fluid is referred to hereinafter as:
\begin{equation}\label{Energy_SPH_Verhagen}
\alpha\,:=- \frac{\Delta E_{fluid}^{dissipation}}{4\,m_{liquid}\,g\,h\,\Phi^{\frac{3}{2}}}\,.
\end{equation}
%
Being $\alpha$ of order unity, the reference energy, $4\,m_{liquid}\,g\,h\,\Phi^{\frac{3}{2}}$ models that part of the mechanical fluid energy (kinetic plus gravitational potential) which is available to be dissipated in breaking.

\subsection{Numerical predictions of the torque exerted by the fluid and the associated dissipation}\label{SPH}
The theoretical model presented in section \ref{sec:Verhagen} is not expected to be valid for large oscillation
amplitudes, as is the case for some used in the present work, or for very small oscillations where hydraulic jumps do not occur.
For this reason, numerical simulations in a 2D framework are performed using the Smoothed Particle Hydrodynamics model
discussed and validated for sloshing flows in \cite{Bouscasse_etal2013} and in \cite{Antuono_etal_JFM2012}.

The filling height adopted is equal to $h=0.092$ m. This choice is motivated by the points discussed in section \ref{sec:fullycoupled}.

\begin{figure}[b!]
\centering
\includegraphics[width=0.48\textwidth]{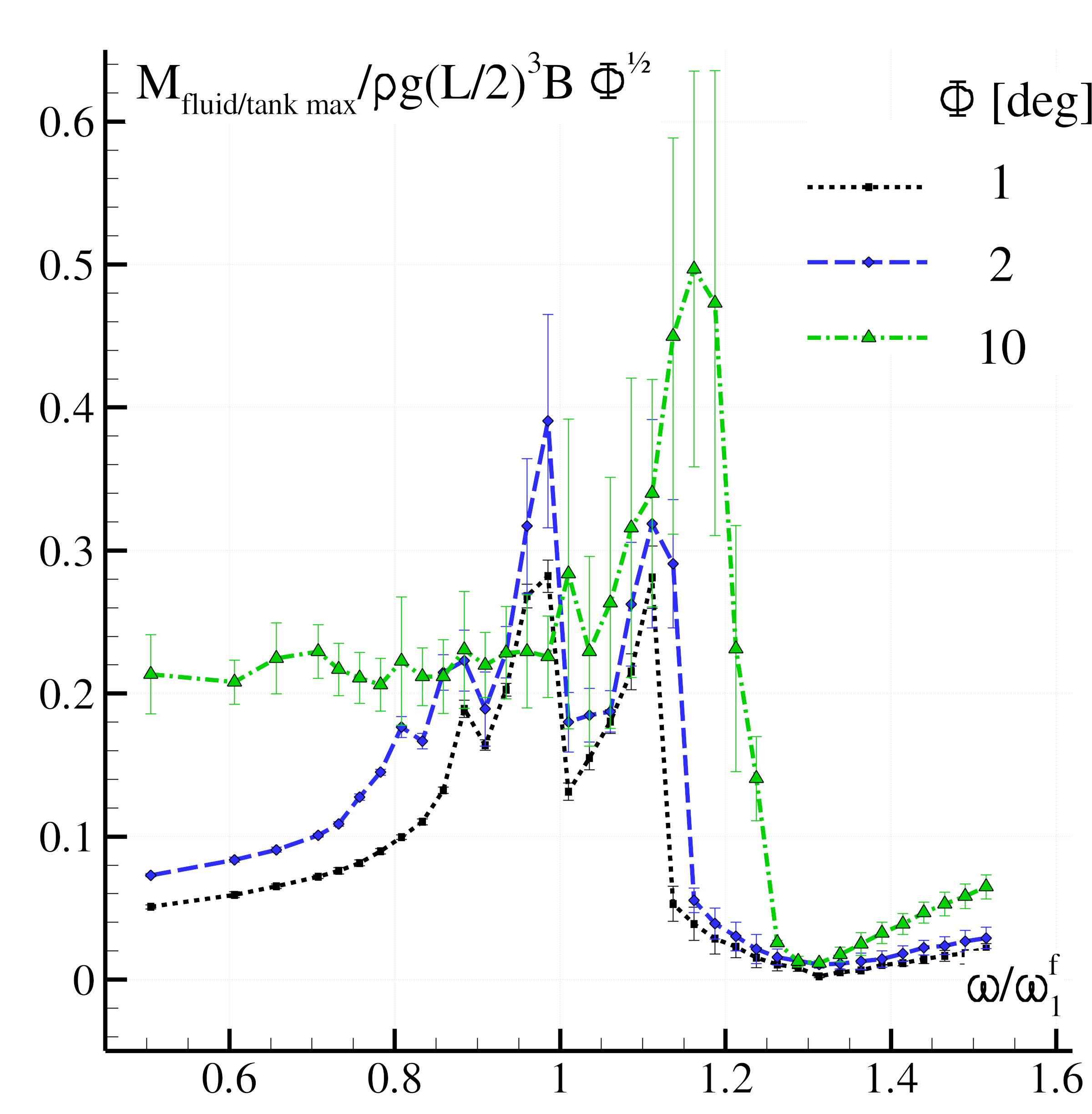}
\includegraphics[width=0.48\textwidth]{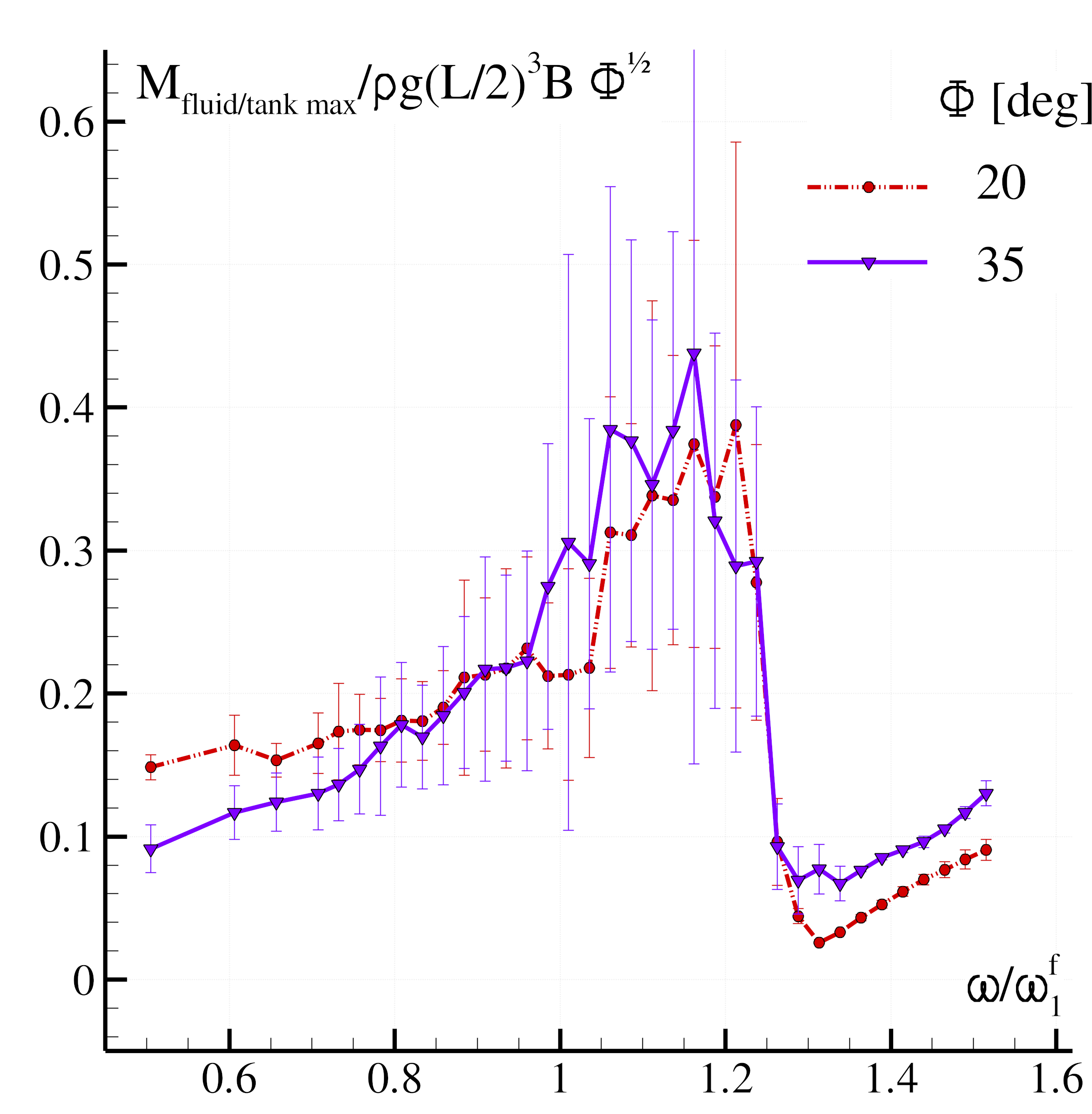}
\caption{
Maximum value for the torque $M_{fluid/tank}$ predicted by the SPH method during the time-periodic state varying the excitation frequency $\omega$. The maximum values plotted are the average of those obtained for the simulated periods. The error bars indicate the associated standard deviation. Left: roll amplitude $\Phi$ = 1,2, 10 degrees. Right: $\Phi$ = 20 and 35 degrees.
}
\label{fig:SPH_Sloshing_Operator}
\vskip 0.2cm
\end{figure}

Plots in Fig. \ref{fig:SPH_Sloshing_Operator} show the maximum torque $M_{fluid/tank}$
recorded in the time-periodic regime for five different roll amplitudes $\Phi$: 1, 2, 10, 20 and 35 degrees
and a range of exciting frequencies $\omega$ close to $\omega^f_1$.

The peak values of $M_{fluid/tank}$ in each oscillation cycle have a very different frequency behavior for
small and large roll angles. Furthermore, for small roll amplitudes, the associated standard deviation across these cycles
is very low. This result indicates repeatability, a characteristic of non-breaking sloshing flows.
For a roll angle greater than $2$ degrees, breaking waves occur, inducing a standard deviation on the
evaluated $M_{fluid/tank}$ that increases with $\Phi$.

The analytical prediction of $M_{fluid/tank}$ in the proximity of $\omega_1^f$
( see equation (\ref{eq:M_Verhagen}) ) is $M_1 \simeq \, 0.457 \sqrt{\Phi}$.
The theoretical model tends to exceed the SPH predictions, however,
the agreement between the theory and the numerics on the maximum torque
remains fair for all investigated roll angles.

\begin{figure}[t!]
\centering
\includegraphics[width=1.0\textwidth]{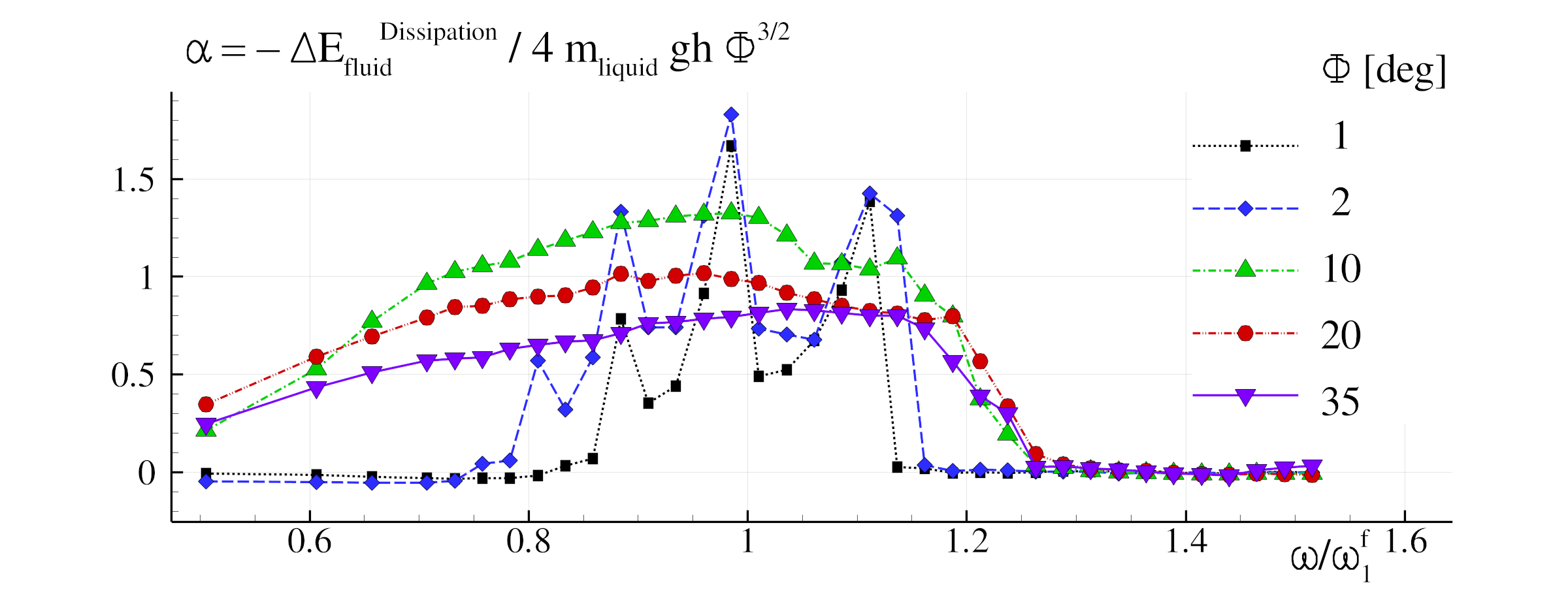}
\caption{
The energy $\Delta E_{fluid}^{dissipation}$ predicted by the SPH method during the time-periodic state is plotted for an excitation frequencies range $(0.4<\omega/\omega_1^f<1.5)$ and for five different roll amplitudes $\Phi$.
}
\label{fig:SPH_Sloshing_Operator_Energy}
\vskip 0.2cm
\end{figure}

Fig. \ref{fig:SPH_Sloshing_Operator_Energy} depicts the $\alpha$ coefficient defined in equation \ref{Energy_SPH_Verhagen} for the five previously
defined roll amplitudes $\Phi$.
For the lowest roll amplitudes, the obtained $\alpha$-values present a complex frequency behaviour with
different peaks linked to secondary resonance effects. Besides this, for $\Phi$ equal to 1 and 2 degrees,
$\alpha$ is very close to the value $1.68$ predicted by the analytical expression(\ref{Energy_Verhagen})
when $\omega$ is close to the frequency $\omega_1^f$.
These results are compatible with those found in \citet{landrini_etal_jfm2007} where the energy dissipated
by breaking waves, when simulating hydraulic jumps with SPH, was shown to be similar to analytical results.

Increasing the roll amplitude, SPH predicts a reduction of the viscous coefficient $\alpha$
which remains in the range of variation $\alpha\in(0.8,1.8)$ for all the five amplitudes studied.
This reduction is also confirmed by the experimental measurements presented in the Part II.
%
\section{The fully coupled angular motion system}\label{sec:fullycoupled}

\subsection{General}\label{generalcoupled}
Once the empty tank mechanical system and the fluid system have been independently analyzed, it is relevant to observe them coupled.

If the fluid is considered ``frozen'', considering a filling height $h=0.092$ m and a ``frozen'' liquid with density $\rho=1000\,Kg/m^3$, the moment of Inertia $I_0$ has an 8\% variation. However,
since $S_g$ also varies, the effect on the mechanical resonance frequency $\omega_1^m$
is limited to a decrease of 0.5\%. Therefore, the effects of the presence of a liquid inside the tank, are mainly due to the induced sloshing flows and not so much to the liquid mass added in the system.

In the previous sections the non-linear empty tank and the forced sloshing dynamics have been described. To analytically study the frequency behavior of the coupled fluid/rig TLD system the methodology described in the works of Frandsen \cite{Frandsen2005} or Alemi Ardakani et al \cite{ardakani2012resonance} should be followed.

In other articles (e.g. Tait's\cite{Tait2008}), the sloshing dynamics of the TLD system is approximated as a simple secondary mass-spring system. This allows (as with the model of Frandsen \cite{Frandsen2005}) the selection of an optimal mass of fluid in order to reduce the oscillation amplitude at the resonance frequency of the mechanical system $\omega_1^m$.

In the case studied here for the pendulum-TLD, the roll motion makes these analytical approaches more complex. Furthermore, for the large range of excitation amplitudes $A_m$ investigated here the linearized approach fails.

The optimal choices for the mass of fluid found in the linearized approaches
are in the neighborhood of $\omega_1^f/\omega_1^m\approx 1$ and this can be explained by the following simple considerations:
\begin{enumerate}
\item
As discussed in section \ref{ss:verhagenmfluid}, if the system is forced at $\omega=\omega_1^f$ then $\Psi_1=-90^\circ$ (the first harmonic of $M_{fluid/tank}$ is in quadrature with the tank motion).
\item
The largest counteraction expected is when $M_{fluid/tank}$ is lagged $180^\circ$ with respect to $M_{mass/tank}$.  Looking at Fig. \ref{fig:momentosyangulos}, this case corresponds to $\delta+\Psi_1=0.$
\item
For the smallest forcing ($A_m=0.05\mathrm{m}$, see section \ref{empty_tank:gen}), the roll motion of the mechanical system with an empty tank is in quadrature ($\delta=90^\circ$) with the sliding mass motion when the system is forced at the mechanical resonance ($\omega=\omega_1^m$).
\item
From the above considerations, when the first sloshing frequency is equal mechanical resonance frequency of the system:
\begin{equation}\label{eq:w1meqw1f}
  \begin{array}{ccc}
  \omega_1 \,:= \omega_1^f\, =\,\omega_1^m  & \qquad \Rightarrow \qquad & \qquad
 g\, \pi/L \tanh(\pi h/L)\,=\,-g\,S_g/I_0,
\end{array}
\end{equation}
which allows identifying the filling height $h=0.092$ m.
\end{enumerate}
Summarizing the above considerations for all the different torques, it is possible to define phasors on a complex plane using the modulus and phases of the first harmonic components obtained by a Fourier decomposition (as in equation \ref{Mfluid_verhagen1}).

\begin{figure}[ht!]
\centering
\includegraphics[width=0.50\textwidth]{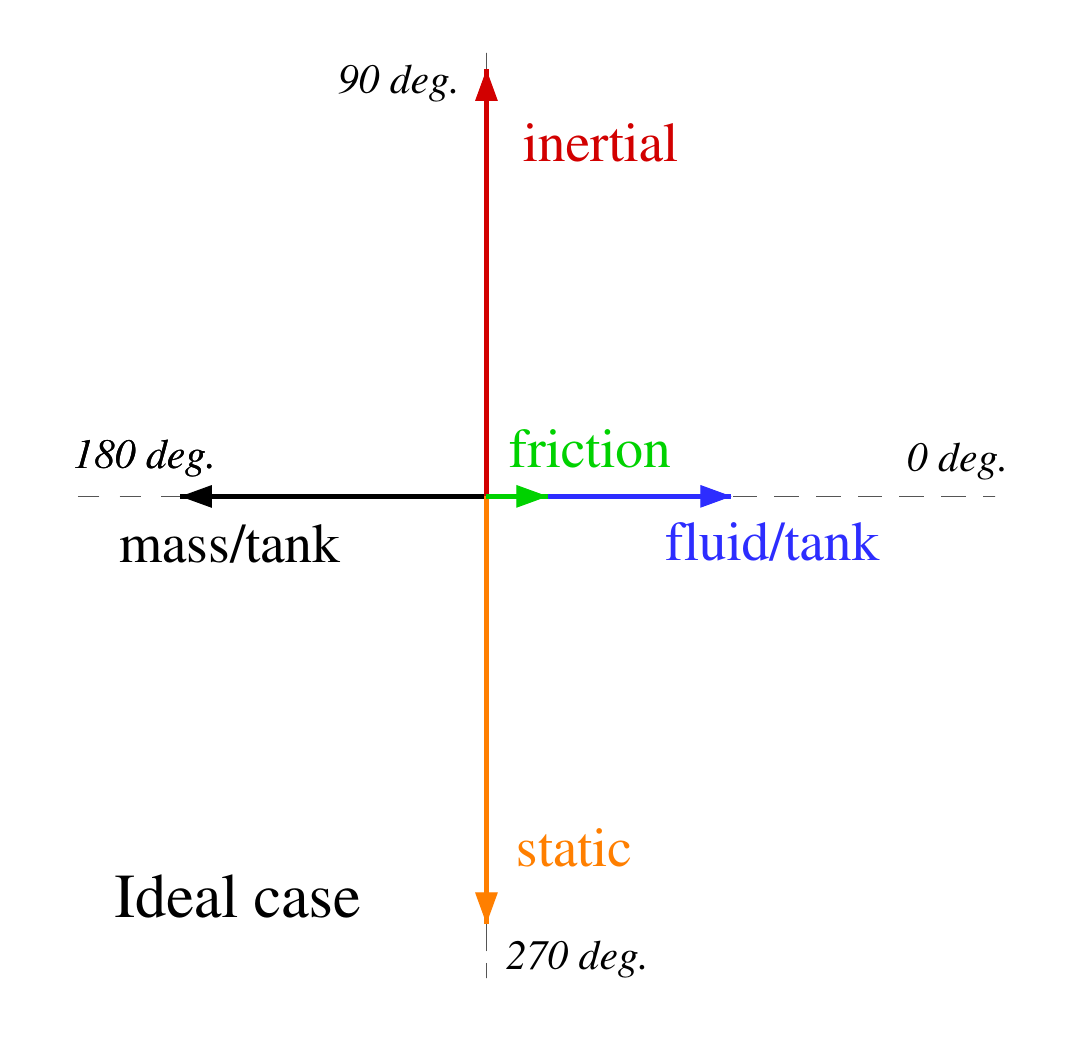}
\caption{Torque modulus and phase. Phasor expected for a tank filled with an inviscid liquid at oscillating with
small amplitude angle, for resonance condition of both liquid and mechanical system.}
\label{fig:Phasor_linear}
\end{figure}

The phasors expected for an idealized system are sketched in Fig. \ref{fig:Phasor_linear}.
The inertial and static components are defined respectively as
$M_{inertial}=-I_0 \ddot{\phi}$ and $M_{static}=g S_g \sin{\phi}$.
The origin of the phases is given by the sliding-mass motion.

In real cases, the dynamic system moves away from this ideal condition.
As a consequence, the optimum choice for the filling height, $h^*$, is not necessarily obtained by (\ref{eq:w1meqw1f}) since $h^*$ also varies due to non-linearities with respect to the forcing amplitude $A_m$ or to the nature of the fluid. These non linearities are the subject of the present study, where the filling height $h$ is therefore set to 0.092 m and this choice is tested across a range of frequencies using a suitable numerical solver to get the fluid reaction $M_{fluid/tank}$.
\subsection{Pendulum TLD: numerical simulation with SPH}
In this section, the Smoothed Particle Hydrodynamics model presented in \cite{Bouscasse_etal2013} and in \cite{Antuono_etal_JFM2012}, is applied to simulate the fully coupled angular motion system. The reader is referred to \cite{Bouscasse2013_floating} for details on the
mechanical system coupling algorithm. The two-dimensional hypothesis is still maintained mainly for computational costs.

\begin{figure}[t!]
\centering
\includegraphics[width=0.45\textwidth]{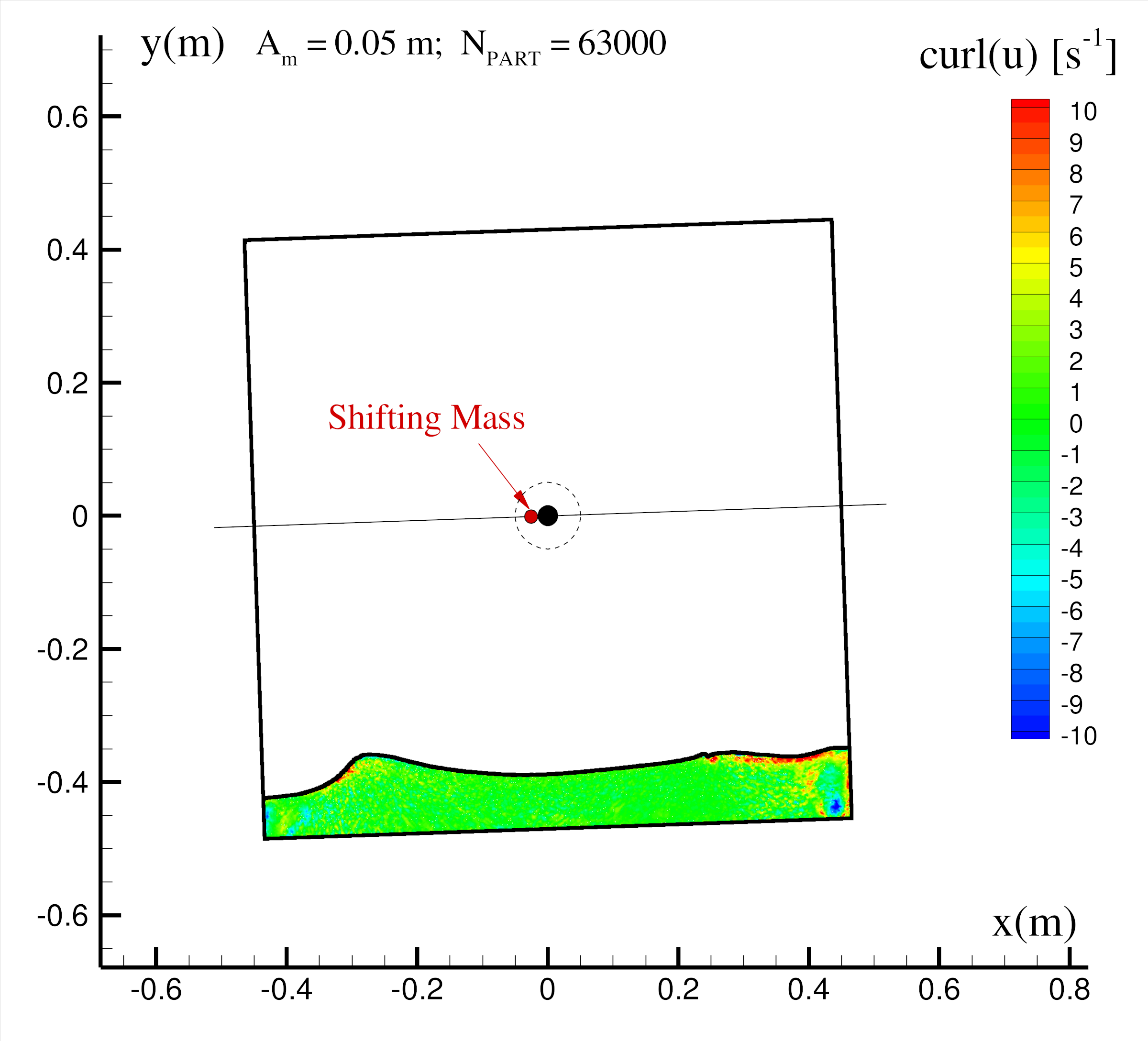}
\includegraphics[width=0.45\textwidth]{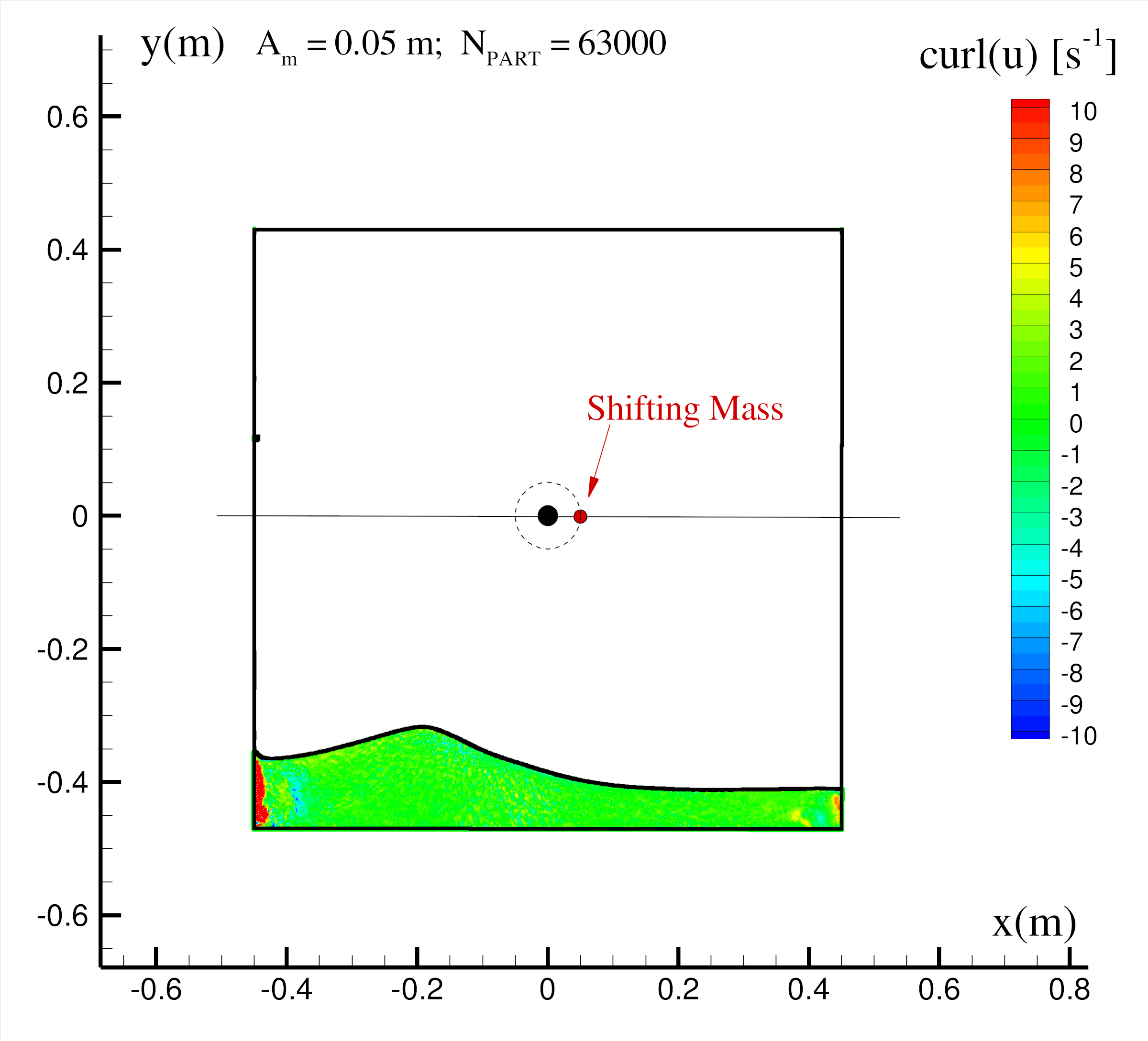}
\includegraphics[width=0.45\textwidth]{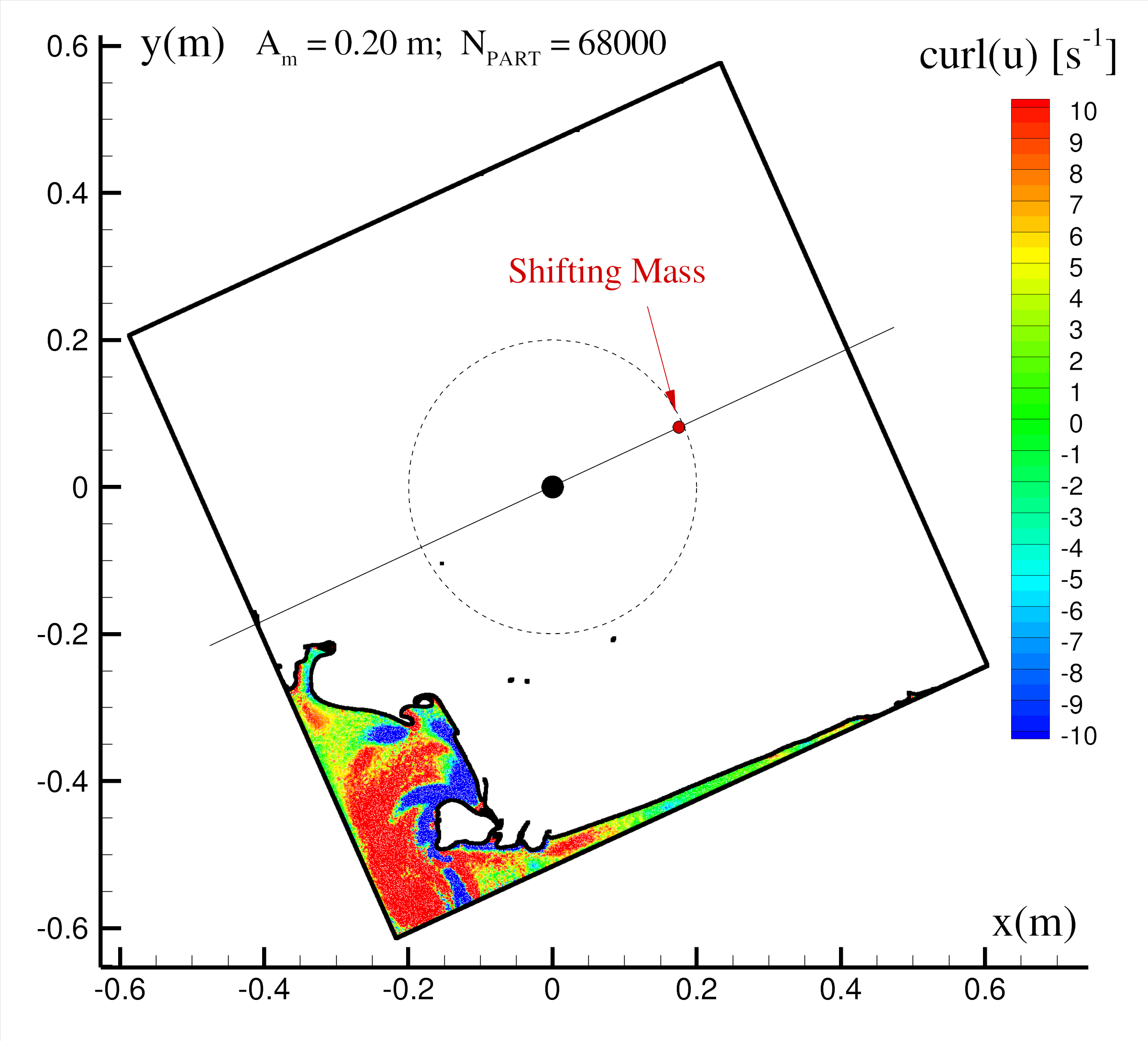}
\includegraphics[width=0.45\textwidth]{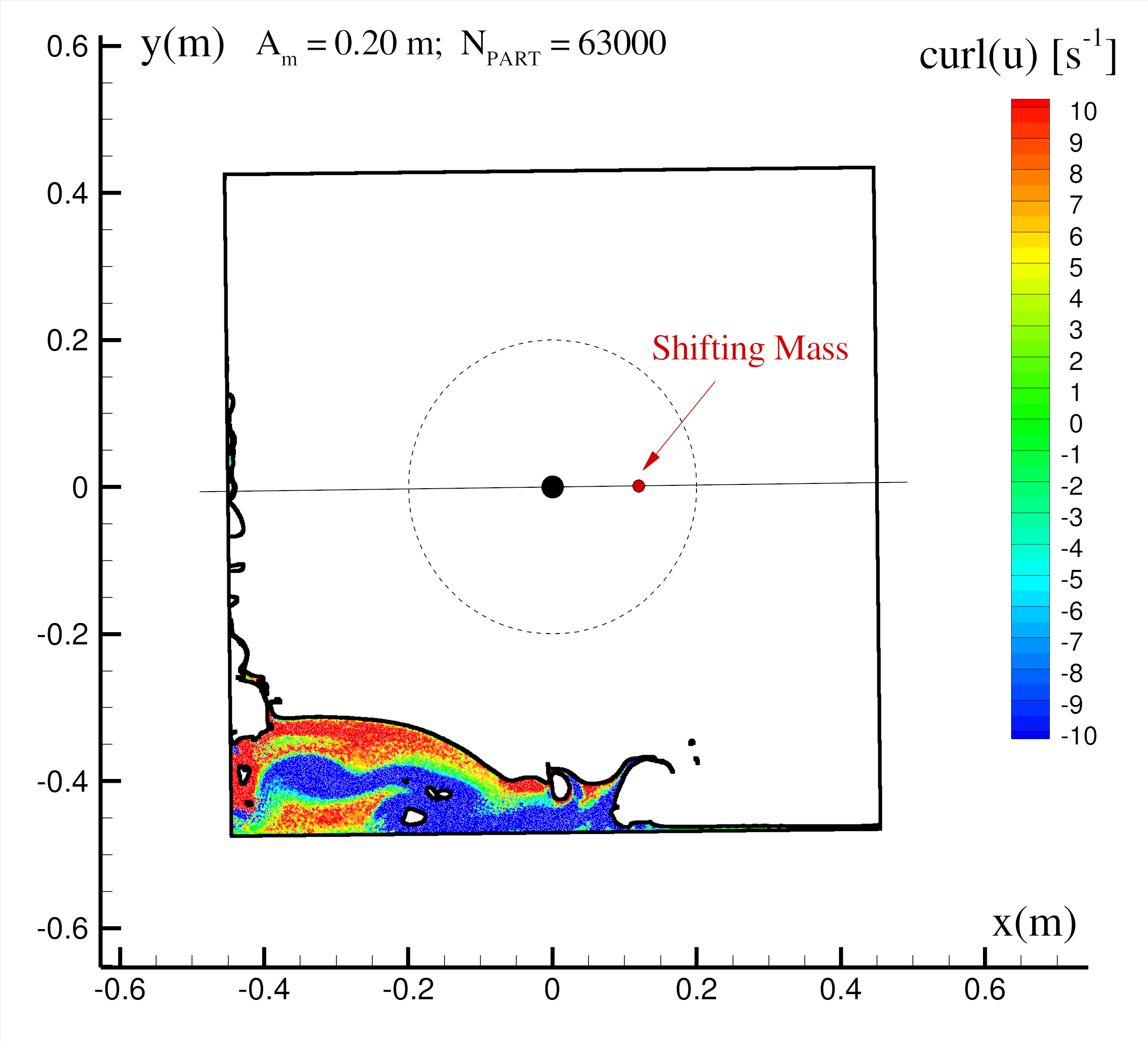}
\caption{Tank filled with water: Sloshing flow predicted by the SPH model
using two excitation amplitudes of the sliding mass: $A_m = 0.05$ m (top) $A_m = 0.20$  m(bottom).
Particles are colored according to their vorticity.}
\label{fig:SPH_FullyCoupled_Snapshot}
\vskip 0.2cm
\end{figure}
Fig. \ref{fig:SPH_FullyCoupled_Snapshot} depicts the sloshing flow predicted by the SPH model.
For the frequency $\omega=\omega_1$, two different excitation amplitudes $A_m$ of the sliding mass are used: $A_m = 0.05$  m and $A_m = 0.20$ m.
For the smallest amplitude, a train wave develops inside the tank and no breaking wave phenomena are predicted;
the motions of the sliding mass and the rolling tank are almost in quadrature. The agreement with the experimental results is very good as it can be seen comparing with the results of Part II.

Conversely, using the highest $A_m=0.20$, the sloshing flow becomes very violent,
with an intense free surface fragmentation process;
for this case, the sliding mass and the roll tank are far away from the quadrature condition.
The SPH prediction of the flow presents no negligible discrepancies with respect to the experiments. Indeed, because of the
violent sloshing condition, the flow comprises air entrapment, turbulence processes and significant three-dimensional effects; they are not modeled by the numerical method.
\begin{figure}[hb!]
\centering
\includegraphics[width=0.8\textwidth]{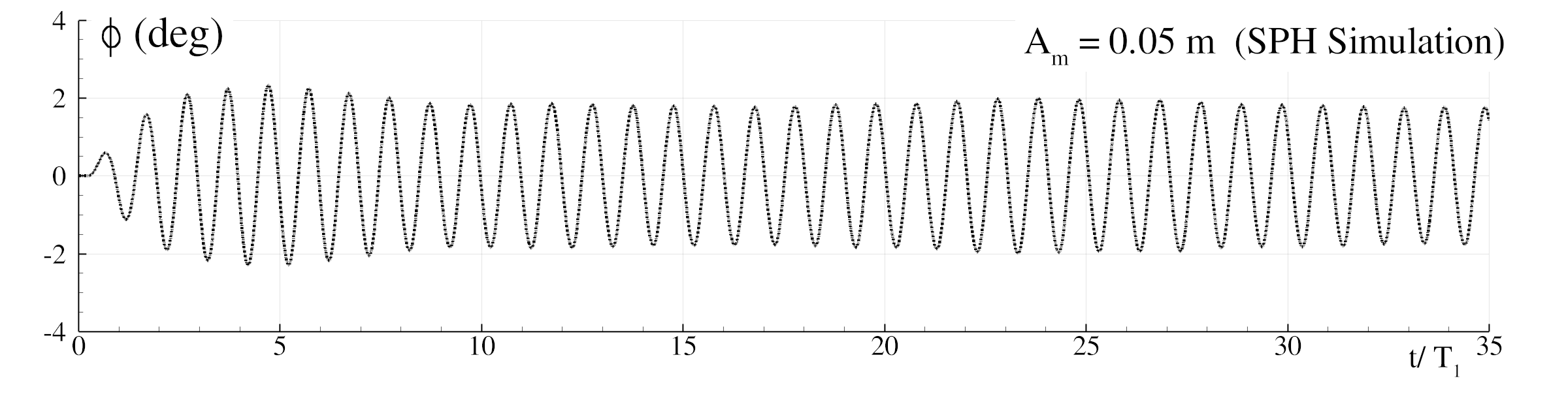}
\includegraphics[width=0.8\textwidth]{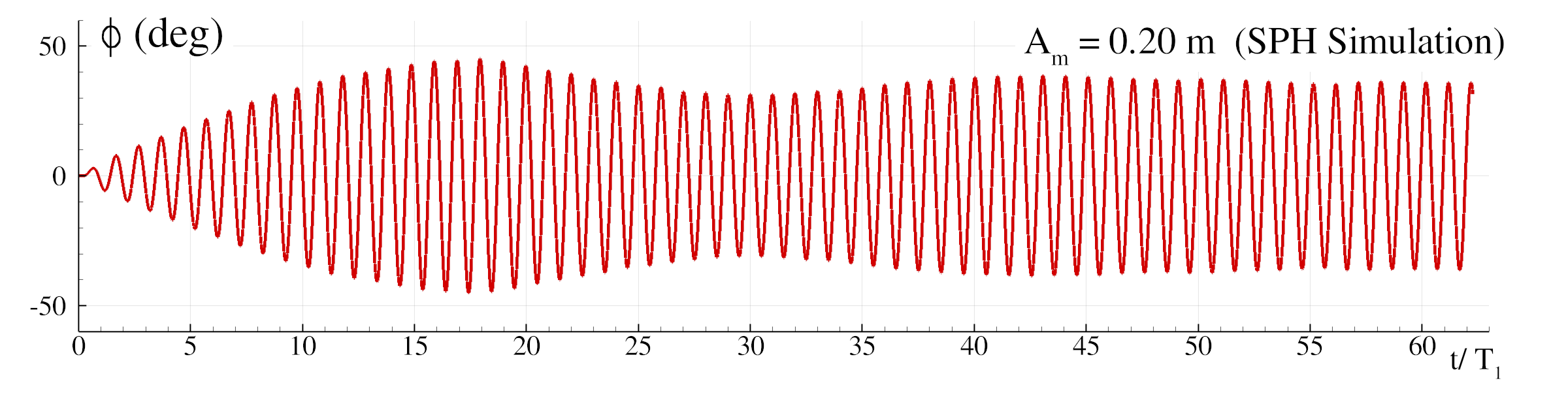}
\caption{Tank filled with water: roll angle plotted as a function of time using $A_m = 0.05$ m (top)
and $A_m = 0.2$ m (bottom)  obtained through a SPH model. }
\label{fig:SPH_Roll_Angle}
\vskip 0.2cm
\end{figure}

For both amplitudes $A_m$, the roll angle and the phase lags predicted by the numerical model agree with the experimental results reported in Part II of this work.
In Fig. \ref{fig:SPH_Roll_Angle} the roll angle $\phi(t)$ predicted by the SPH is plotted as a function of time. For $A_m=0.05$ m an almost time-periodic state is reached after almost ten periods. Even if a small sub-harmonic develops, the roll-angle amplitude $\Phi$ stabilizes at a value of around 2 degrees.
The largest $A_m$ requires more periods of oscillation to reach a a time-periodic state, for which
a value of 35 degrees is attained.
Comparing the maximum roll angles with the ones evaluated with the empty tank
conditions in the case with $A_m=0.05$ m (see section \ref{empty_tank}),
the presence of liquid induces a drastic reduction of the roll motion and the system
thus behaves like a classical TLD.

\begin{figure}[ht!]
\centering
\includegraphics[width=0.7\textwidth]{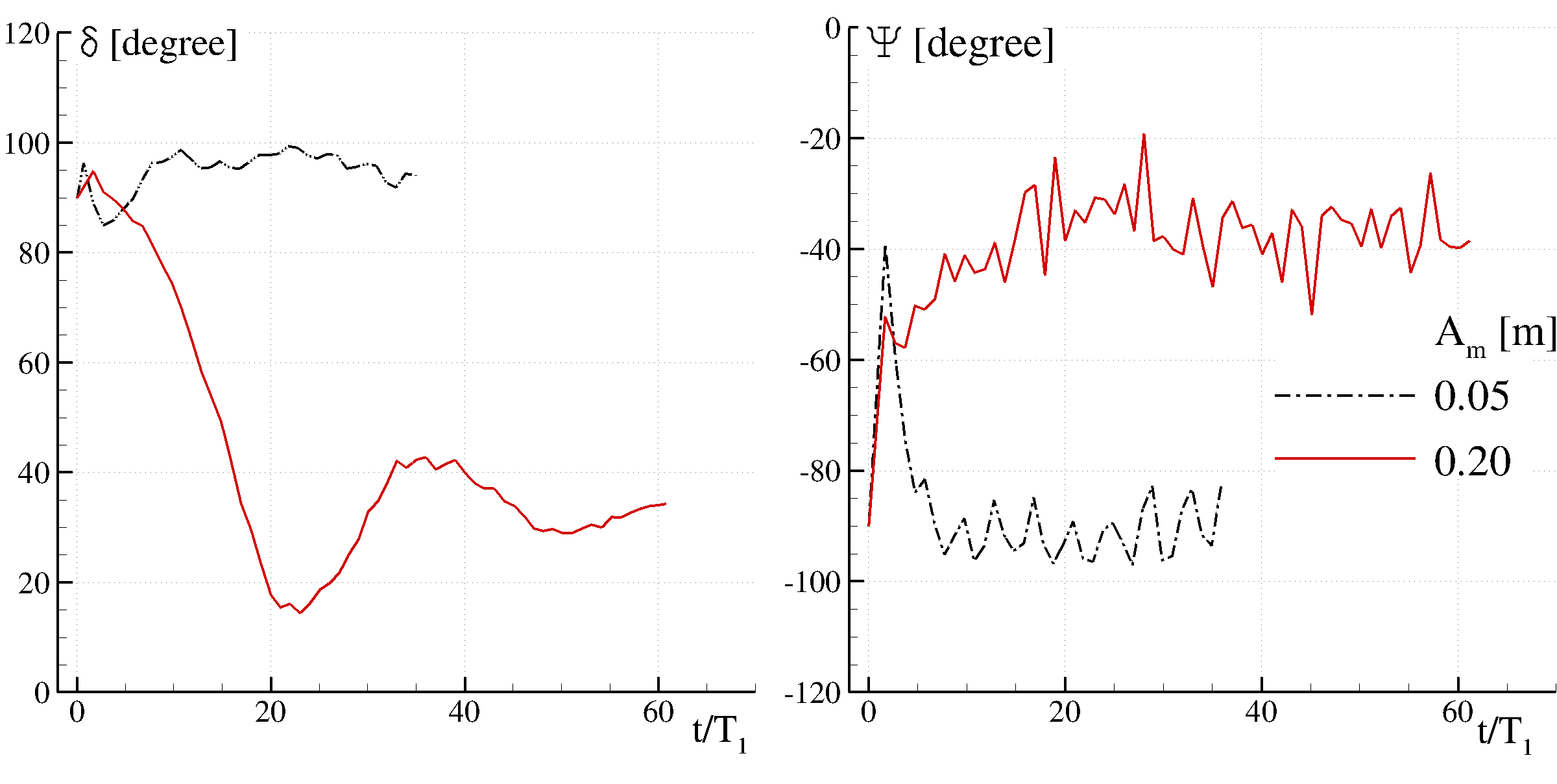}
\caption{Tank filled with water: phase lags $\delta$ and $\Psi$ are plotted as a function of time using four different excitation amplitudes $A_m$ obtained through a SPH model.}
\label{fig:SPH_Delta_Psi}
\vskip 0.2cm
\end{figure}
This is not the case for the largest amplitude $A_m=0.20$ m. Indeed, in such a condition the final roll angle
with water inside the tank is practically the same as obtained with the empty tank condition.
Fig. \ref{fig:SPH_Roll_Angle} shows the phase lags $\delta$ and $\Psi$ predicted by SPH and plotted as a function of time.
Since the roll motion $\phi(t)$ is not affected by super-harmonics, $\delta$ presents a smooth time behavior.
Conversely, $\Psi(t)$ displays a noisy time history, which is linked to the more complex numerical treatment of $M_{fluid/tank}$ (see \ref{sec:Preliminary_Consideration})

For $A_m=0.05\;\mathrm{m}$, $\delta$ and $\Psi$ are close to 90 degrees and -90 degrees respectively. The system
is therefore close to the ideal condition discussed in section \ref{generalcoupled}.
For $A_m=0.20$ m the time history of $\delta$ is more complicated and only begins to stabilize after 60 periods of oscillation at around 35 degrees. This condition gives an indication of how the non-linearities of the dynamical system play a relevant role for this second case.
This unique behavior will be discussed in greater detail in Part II of the manuscript.
%
%
%
%
%
\subsection{Pendulum TLD: frequency behavior}
Since the SPH model seems to predict the time evolution of the coupled system with sufficient accuracy, it is also used
to study the frequency behavior.

\begin{figure}[ht!]
\centering
\vskip 0.2cm
\includegraphics[width=0.8\textwidth]{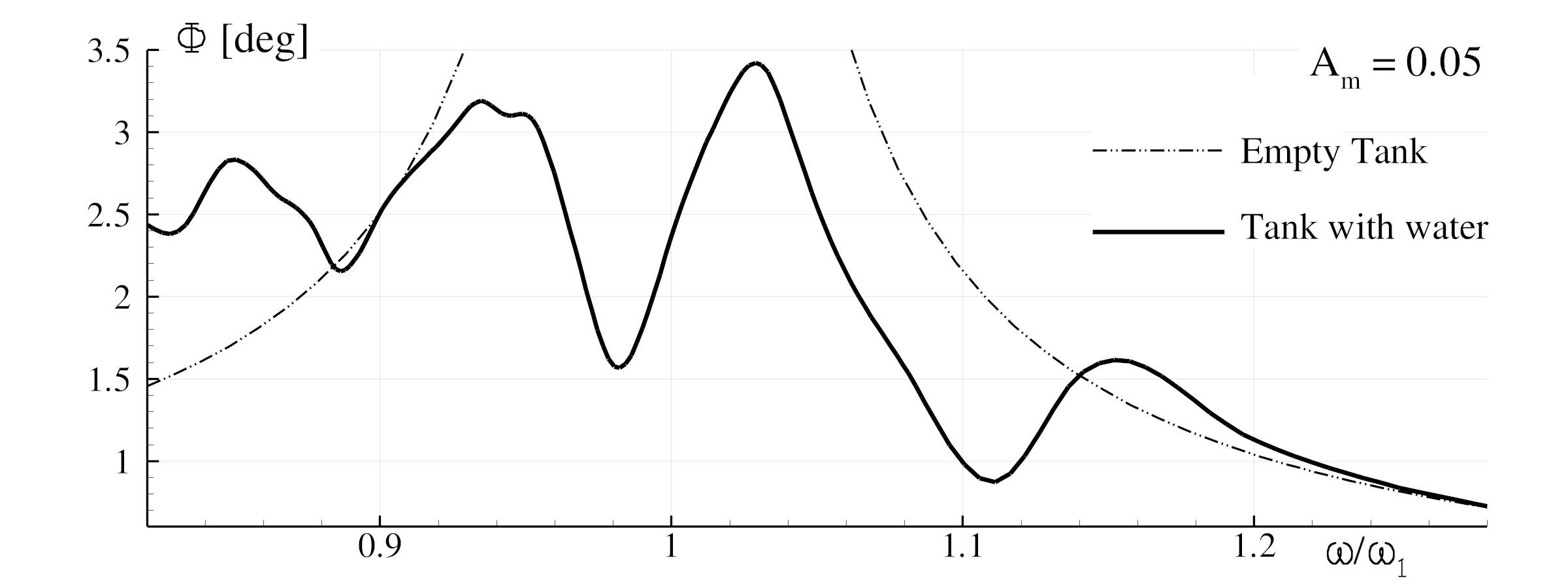}
\caption{Fully coupled angular motion system.
Frequency operators evaluated through an SPH model, for Roll angle $\Phi$, reached at periodic state for the excitation amplitude $A_m$= 0.05 m}
\label{fig:SPH Operator_Coupled_zoom}
\end{figure}

For the smallest amplitude ($A_m=0.05\;\mathrm{m}$, Fig. \ref{fig:SPH Operator_Coupled_zoom} shows the roll angle $\Phi$ reached at time-periodic state for a range of frequencies. $\Phi(\omega)$ presents four peaks while the analysis proposed in \cite{Frandsen2005} shows a classical TLD system which presents only two peaks around the mechanical resonant condition.
However, the reduction of $\Phi$ in the neighborhood of $\omega_1$, is also maintained in the present system.

\begin{figure}[ht!]
\centering
\vskip 0.2cm
\includegraphics[width=0.9\textwidth]{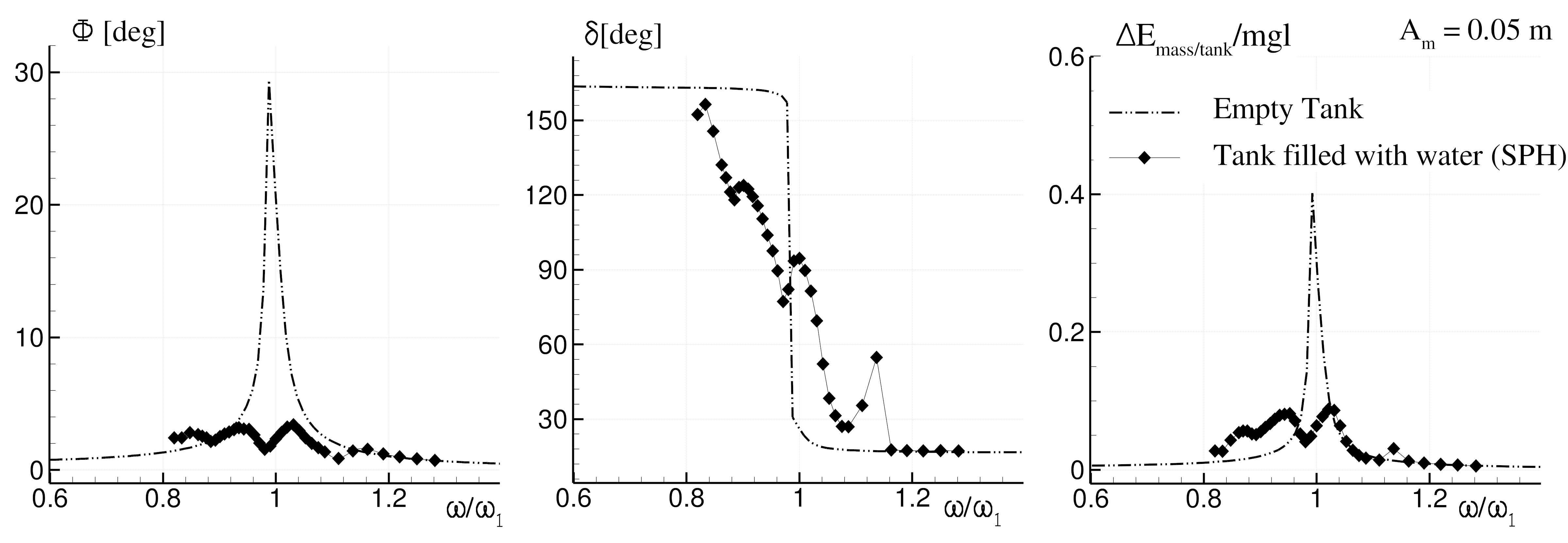}
\includegraphics[width=0.9\textwidth]{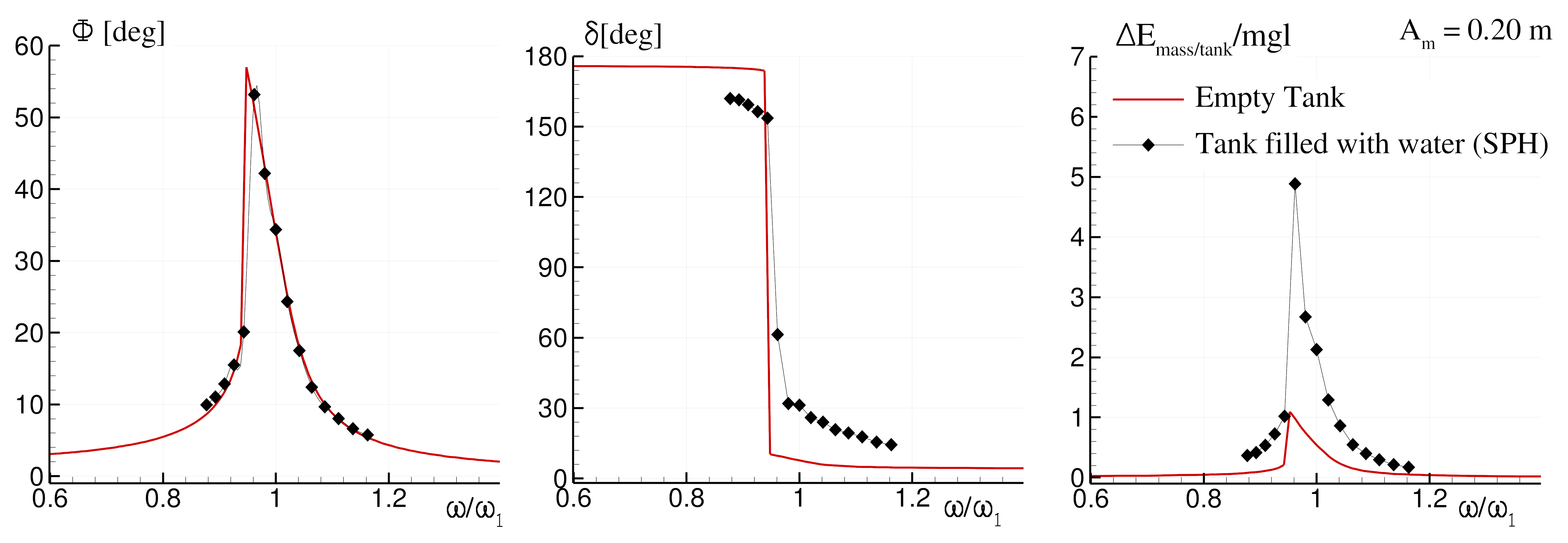}
\caption{Fully coupled angular motion system.
Frequency operators evaluated through an SPH model, for the Roll angle $\Phi$, the phase lag $\delta$ and energy transfer between the moving mass and the tank, $\Delta E_{mass/tank}$, reached at periodic state for two different excitation amplitude $A_m$: 0.05 m (top panel), 0.20 m (bottom panel).}
\label{fig:SPH Operator_Coupled}
\end{figure}

Fig. \ref{fig:SPH Operator_Coupled} shows the frequency operators on the roll angle $\Phi$, the phase lag $\delta$ and the energy transfer between the moving mass and the tank, $\Delta E_{mass/tank}$, reached at time-periodic state for two different excitation amplitudes $(A_m=0.05,\, 0.20\, \mathrm{m})$. The operators for the empty tank condition presented in section \ref{empty_tank:gen} are reported in this plot to highlight the differences induced by the sloshing liquid.

For $A_m=0.05$ m the frequency behavior of the phase lag $\delta$ is very complex as a result of the shallow water sloshing dynamics. However, for $\omega=\omega_1$ the coupled system confirms that it is close to a quadrature condition.
Since the rolling motion is highly reduced in the presence of water, the work done by the sliding mass  $\Delta E_{mass/tank}$ is smaller than in the empty tank condition.

Conversely, for $A_m$ equal to 0.20 m, the $\Phi$ angles reached at time-periodic state are practically not affected by the presence of the water,
and although there are some visible effects on the phase lag $\delta$, the main differences appear on $\Delta E_{mass/tank}$.
Indeed, when water is present inside the tank and for a large $A_m$, the sloshing flow is not able to reduce the roll motion and the system does not perform as an efficient TLD.  However, the work exerted by the sliding mass, $\Delta E_{mass/tank}$, increases up to a factor of five.
This phenomenon will be further described in Part II.

%
\section{Conclusions}

The kinematics, dynamics and energy dissipation mechanisms of a pendulum-TLD system have been analyzed.

The pendulum-TLD is composed of three coupled sub-systems: first, a sliding mass whose weight excites the motion, second, the moving parts, including the empty tank, of an angular motion sloshing rig, and third, the fluid which partially fills that tank.

An analogy with TLD and HMLD systems has been provided. Differently from other TLDs studied in the literature, the Pendulum-TLD involves large motions and complex flows, which do not permit the use of an analytical fluid dynamic model.

The nonlinear dynamics of the Pendulum-TLD has been documented
both for the empty tank and for the tank partially filled with water.
The frequency behavior of the roll angles, phase lags and energy transfer has been discussed.

The energy dissipated by the sloshing flow has been quantified through a simple theoretical model based on hydraulic jump solutions. This model allows for an evaluation of the mechanical energy available to be dissipated in breaking.
Furthermore, a scaling factor for the energy available to be dissipated in breaking has been obtained from this analysis. This scaling factor has been used to make non dimensional in a meaningful way the numerical results obtained by an SPH model (and the experimental data of part II).

From the numerical simulations, the complex kinematics and dynamics of the flow has been discussed: low amplitude traveling waves occur for the small excitation cases while breaking waves and violent fluid-structure impacts develop for large excitations.

Through the SPH model the complete frequency behavior of the fully coupled system has been obtained. Interesting features have been identified. Specifically, for small excitations, the system behaves like a classical TLD. The frequency response changes drastically with large excitations.

The present work is completed with the experimental analysis conducted in part II of this paper series.

\section*{Acknowledgements}
The research leading to these results has received funding from
the Spanish Ministry for Science and Innovation under
grant TRA2010-16988 ``\textit{Ca\-rac\-te\-ri\-za\-ci\'on Num\'erica y
Experimental de las Cargas Fluido-Din\'amicas en el transporte de Gas Licuado}'' .

This work has been also funded by the Flagship Project RITMARE - The Italian Research for
the Sea - coordinated by the Italian National Research Council and funded by the Italian Ministry of
Education, University and Research within the National Research Program 2011-2013.

The authors thanks the reviewers for the useful suggestions.

The authors are grateful to Sonny Mendez and Hugo Gee for English language proofreading.
\bibliographystyle{apsrev}
\bibliography{bib}
\end{document}